\documentclass[preprint,review,12pt]{elsarticle}
\usepackage[margin=1.0in]{geometry}

\usepackage{graphicx,color}
\usepackage{dcolumn}
\usepackage{bm}
\usepackage{amsmath,amsfonts,bbm}
\usepackage{amsthm}
\usepackage{multirow}
\usepackage{subfigure}
\usepackage[rightcaption]{sidecap}
\usepackage{array}

\def\mapright#1{\buildrel{#1}\over\longrightarrow}

\def\undermapright#1{\mathrel{\mathop{\kern0pt \longrightarrow}\limits_{#1}}}
\def\undermapleft#1{\mathrel{\mathop{\kern0pt \longleftarrow}\limits_{#1}}}
\def\maprightleft#1#2{\vcenter{\offinterlineskip \hbox{$\displaystyle
  \mapright{#1}$}\kern-1pt\hbox{$\displaystyle\undermapleft{#2}$}}}

\def\E{\mathbb{E}}
\def\Var{\mathtt{Var}}

\def\x{\mathbb{x}}
\def\P{\mathcal{P}}

\def\f{\frac}

\newcolumntype{C}[1]{>{\centering\let\newline\\\arraybackslash\hspace{0pt}}m{#1}}

\newtheorem{theorem}{Theorem}[section]
\newtheorem{lemma}[theorem]{Lemma}
\newtheorem{definition}[theorem]{Definition}

\journal{Computers \& Mathematics with Applications}

\begin{document}

\begin{frontmatter}



\title{Implicit Simulation Methods for Stochastic Chemical Kinetics}


\author[a]{Tae-Hyuk~Ahn\corref{cor1}}
\ead{ahnt@ornl.gov} \cortext[cor1]{Corresponding authors.}
\author[b]{Adrian~Sandu\corref{cor1}}
\ead{sandu@cs.vt.edu}
\author[c]{Xiaoying Han}
\ead{xzh0003@auburn.edu}

\address[a]{Computer Science and Mathematics Division, Oak Ridge National Laboratory, Oak Ridge, TN 37831, USA}
\address[b]{Computational Science Laboratory, Department of Computer Science, Virginia Polytechnic Institute and State University, Blacksburg, VA. 24061, USA}
\address[c]{Department of Mathematics and Statistics, Auburn University, Auburn, AL. 36849, USA}

\begin{abstract}
In biochemical systems some of the chemical species are present with
only small numbers of molecules. In this situation discrete and
stochastic simulation approaches are more relevant than continuous
and deterministic ones. The fundamental Gillespie's stochastic
simulation algorithm (SSA) accounts for every reaction event, which
occurs with a probability determined by the configuration of the
system. This approach requires a considerable computational effort
for models with many reaction channels and chemical species. In
order to improve efficiency, tau-leaping methods represent multiple
firings of each reaction during a simulation step by Poisson random
variables. For stiff systems the {\em mean} of this variable is
treated implicitly in order to ensure numerical stability.

This paper develops fully implicit tau-leaping-like algorithms that
treat implicitly {\em both the mean and the variance} of the Poisson
variables. The construction is based on adapting weakly convergent
discretizations of stochastic differential equations to stochastic
chemical kinetic systems. Theoretical analyses of accuracy and
stability of the new methods are performed on a standard test
problem. Numerical results demonstrate the performance of the
proposed tau-leaping methods.
\end{abstract}


\end{frontmatter}


\section{Introduction} \label{sec:intro}
Biological systems are frequently modeled as networks of interacting
chemical reactions. In systems formed by living cells  stochastic
effects are very important, as typically some reactions involve only
a small number of molecules (of one or more species)
\cite{McAdams97}. The \emph{Chemical Master Equation}
(CME)~\cite{Gill92, Kampen07} governs the time-evolution of the
probability function of the system's state. Gillespie proposed the
stochastic simulation algorithm (SSA), a Monte Carlo approach based
on sampling exactly the probability density evolved by the
CME~\cite{Gill77}.  Since each reaction is accounted for
individually, the overall computational effort becomes an issue with
systems of practical interest. This motivates the development of
approximate sampling algorithms that trade some accuracy in order to
considerably improve computational efficiency.

One approximate acceleration procedure is the ``tau-leaping
method''~\cite{Gill01}, in which multiple reactions are simulated
within a pre-selected time interval of length $\tau$. The
tau-leaping method requires that $\tau$ satisfies the ``leap
condition'': the expected state change induced by the leap must be
sufficiently small such that propensity functions remain nearly
constant during the time step $\tau$.  In this case the number of
times that each reaction fires in the interval $\tau$ is
approximated by a \emph{Poisson random variable}.

While the tau-leaping method is efficient for single timescale
systems, it becomes unstable for stiff systems when the stepsize
$\tau$ is large. Stiffness characterizes the dynamics where
well-separated ``fast'' and ``slow'' time scales are present, and
the ``fast modes'' are stable. The implicit tau-leaping method
improves the numerical stability~\cite{Rath03}, but it has a damping
effect and its results have much smaller variances than SSA results.
The trapezoidal tau-leaping formula was proposed to reduce this
damping effect~\cite{Cao05}. Additional approaches have been
developed to accelerate the efficiency of the exact SSA through
various approximations~\cite{Cao04, Gill03, Rath05}. Improved step
size ($\tau$) selection is discussed in~\cite{Gill01, Gill03}.
An alternative point of view is to understand the tau-leaping method
as the Euler scheme for stochastic differential equations
(SDEs)~\cite{Tian01, Li07, Hu11}, applied to stochastic chemical
kinetics. This is the point of view taken in this paper. We propose
new tau-leaping-like methods motivated by weakly convergent discrete
time approximations of  stochastic differential
equations~\cite{Kloeden99}.

The existing implicit tau-leaping methods treat implicitly only the
mean part of the Poisson variables; the variance part is treated
explicitly. Therefore current algorithms can be characterized as
partially implicit. This paper develops several fully implicit
algorithms, where both the mean and the variance parts of the random
variables are solved implicitly. The ``BE--BE'' method uses the
stochastic backward Euler method for both the mean part and the
variance part of the Poisson variables. The ``BE--TR'' method uses
the implicit stochastic trapezoidal method for the variance part of
the Poisson variables. The ``TR--TR'' method discretizes both the
mean and the variance of the Poisson variables with the trapezoidal
method. This work also proposes implicit second order weak Taylor
tau-leaping methods for the stochastic simulation of chemical
kinetics. Numerical stability is investigated theoretically in the
context of the reversible isomerization reaction test problem, an
approach that is well accepted
 \cite{Cao04-2, Hu11}.

Numerical experiments are performed with three different chemical
systems to assess the efficiency and accuracy of the new implicit
algorithms. The numerical results show that the proposed methods are
accurate, with an efficiency comparable to that of the original
implicit tau-leaping methods. They confirm the theoretical stability
analysis conclusions that out of the six new methods four are
unconditionally stable, and two are conditionally stable. These
analyses perfectly explain our preliminary results reported
previously \cite{Ahn11-HPC11, Ahn11-ICCS11}. The numerical
experiments show that, for stiff systems, all three fully implicit
tau-leaping methods avoid large damping effects and are stable for
any stepsize \cite{Ahn11-HPC11}. But two of the implicit second
order weak Taylor methods show unstable behavior for large stepsizes
(although they are more stable than the explicit tau-leaping method
\cite{Ahn11-HPC11}).

The remaining part of the paper is organized as follows. Section
\ref{sec:SSA} describes the traditional SSA algorithm. Numerical
schemes for the solution of SDEs are presented in Section
\ref{sec:SDEs}. In Section \ref{sec:newSSA} the proposed new methods
are introduced. Section \ref{sec:analysis} performs a numerical
stability analysis using a traditional test example. Results from
numerical experiments with three different systems are presented in
Section \ref{sec:results}. Section \ref{sec:conclusions} draws
conclusions and points to future work.

\section{Stochastic Simulation Algorithms for Chemical Kinetics}
\label{sec:SSA}
In this section we briefly review the traditional SSA and
tau-leaping algorithms for stochastic chemical kinetics.

\subsection{Exact Stochastic Simulation Algorithm}
Consider a biochemical system involving $N$ molecular species $S_1$,
$\ldots$, $S_N$, composed of $M$ reaction channels $R_1$, $\ldots$,
$R_M$.  Denote by $X_i(t)$ the number of molecules of species $S_i$
at time $t$. We are interested to generate the evolution of the
state vector $X(t)=(X_1(t),...,X_N(t))$ starting from an initial
state vector $X(t_0)$.   Assume that the system is well-stirred in a
constant volume $\Omega$ and is in thermal equilibrium at some
constant temperature.  The state change vector $\nu_j = \nu_{\cdot,
j}=(\nu_{1,j}, ..., \nu_{N,j})$ for the channel $R_j$ is defined as
the change in the population of molecule $S_i$ caused by one $R_j$
reaction. The propensity function $a_j$ gives the probability
$a_j(x)dt$ that one $R_j$ reaction will occur in the next
infinitesimal time interval $[t, t+dt)$.

The SSA simulates every reaction event~\cite{Gill77}. With $X(t) =
x$, $p(\tau ,j|x, t)d \tau$ is defined as the probability that the
next reaction in the system will occur in the infinitesimal time
interval $[t+\tau, t+\tau+d\tau)$, and will be an $R_j$ reaction. By
letting $a_0(x)\equiv\sum_{j=1}^M a_j(x)$, the equation
\begin{equation*}
  p(\tau ,j|x, t) = a_j(x) \exp(-a_0(x)\tau)
\end{equation*}
can be obtained. A Monte Carlo method is used to generate $\tau$ and
$j$. On each step of the SSA, two random numbers $r_1$ and $r_2$ are
generated from the uniform (0,1) distribution. From probability
theory, the time for the next reaction to occur is given by
$t+\tau$, where
\begin{equation*}
  \tau={1 \over a_0({x})} \ln \left(1 \over r_1\right).
\end{equation*}
The next reaction index $j$ is given by the smallest integer
satisfying
\begin{equation*}
  \sum_{j'=1}^{j}a_{j'}(x) > r_2\,a_0(x).
\end{equation*}
After $\tau$ and $j$ are obtained, the system states are updated by
$X(t+\tau):=x+\nu_j$, and the time is updated by $t:=t+\tau$. This
simulation iteration proceeds until the time $t$ reaches the final
time.

\subsection{Tau-Leaping Method}
The SSA is an exact stochastic method for chemical reactions,
however, it is very slow for many real systems because the SSA
simulates only one reaction at one time. One of the approximate
simulation approach is the tau-leaping method~\cite{Gill01}. The
basic idea of the tau-leaping method is that multiple reactions can
be simulated at each step with a preselected time $\tau$. The
tau-leaping method requires that the selected $\tau$ must be small
enough to satisfy the leap condition, i.e., the expected state
change induced by the leap must be sufficiently small so that
propensity functions remain nearly constant during the time step
$\tau$.

Given $X(t)=x$, denote by $K_j(\tau;x,t)$ the number of times that
reaction channel $R_j$ fires during the time interval $[t,t+\tau)$
where $j = 1, \ldots, M$. The state $X(t)=x$ is updated by
\begin{equation}
  \label{eq:exp-tau-leap}
  X(t+\tau)=x+\sum_{j=1}^M \nu_j\, K_j(\tau;x,t).
\end{equation}
If the leap condition is satisfied, $K_j(\tau;x,t)$ can be modeled
by a Poisson random variable which counts the number of occurrence
during a given time period.  A Poisson variable with parameter $a$
(denoted by $\P(a)$), takes the value $k$ with a probability
$\P(X=k)=[e^{-a}(a)^k]/k!$. For stochastic chemical systems
$\P(a\tau)$ is interpreted physically as the number of events that
will occur in any finite time $\tau$, given that the probability of
an event occurring in any future infinitesimal time $dt$ is $a\,dt$.
Tau-leaping methods use the approximation
\[
  K_j(\tau;x,t) \approx \P_j(a_j(x)\tau),
\]
where $\P_j$ is a Poisson random variate parameter $a_j(x)\tau$.

\subsection{Implicit Tau-Leaping and Trapezoidal Methods}
In general, the tau-leaping methods are only able to perform well if
they continue to take time steps that are of single timescale as
fast or slow mode. This drawback is caused by the fact that explicit
methods advance the solution from one time to the next by
approximating the slope of the solution curve at or near the
beginning of the time interval.   For a ``stiff'' system with widely
varying dynamic modes among which the fastest mode is stable, the
leap condition is used to bound the step size $\tau$ to be within
the timescale of the fastest mode.  Therefore, large leaps are not
feasible for stiff systems as they result in no advantage compared
to the exact SSA. In addition, forced big time step size $\tau$
might lead to unstable population states.

The tau-leaping method is explicit because the future random state
$X(t+\tau)$ is driven only by an explicit function of current state
$X(t)$. An implicit tau-leaping method~\cite{Rath03} modifies the
explicit tau-leaping method as follows. $\P_j$ can be split as
\begin{equation*}
  \P_j=a_j\tau + (\P_j-a_j\tau).
\end{equation*}
We then evaluate the mean value part $a_j\tau$ and the zero-mean
random part (variance of the Poisson variables) $\P_j - a_j \tau$ at
the known state $X(t)$. Therefore,
\begin{equation}
  \label{eq:imp-tau-leap}
  X(t+\tau)=x+\sum_{j=1}^M \nu_j\left\{ \tau a_j \left(X(t+\tau)\right)
  + \P_j(a_j(x)\tau)-\tau a_j(x)\right\}.
\end{equation}

The implicit equation is solved by Newton's iteration method, and
the floating point state $X(t+\tau)$ is rounded to the nearest
integer values. This implicit tau-leaping method allows much bigger
step size than the explicit tau-leaping method for stiff systems.
But large step sizes might provoke damping effect, which means that
when a large step size is used to solve a stiff system, it yields a
much smaller variance and damps out the natural fluctuations of the
stochastic nature~\cite{Rath03}.

The trapezoidal tau-leaping formula was proposed to reduce the
damping effect of the implicit tau-leaping formula~\cite{Cao05}. The
formula is
\begin{equation}
  \label{eq:imp-trape-tau-leap}
  X(t+\tau)=x+\sum_{j=1}^M \nu_j\left\{{\tau \over 2}
  a_j\bigl(X(t+\tau)\bigr) + \P_j(a_j(x)\tau)-{\tau \over 2}
  a_j(x)\right\}.
\end{equation}
Because the trapezoidal rule has a second order convergence without
damping effect, this formula has better accuracy and stiff stability
than the implicit tau-leaping method. The trapezoidal method,
however, is only second order for the mean value, and still first
order for the variance.

\section{Discrete Time Approximations for SDEs}
\label{sec:SDEs}
This section discusses the numerical solution of stochastic
differential equations (SDEs), with an emphasis on weak
approximations~\cite{Kloeden99}.

\subsection{Stochastic Differential Equations (SDEs)}
SDEs are differential equations that incorporate white noise (the
``derivative'' of a Wiener process) and their solutions are random
processes. Consider the following $d$-dimensional SDE
system~\cite{Kloeden99}
\begin{equation}
  \label{eq:ito-form}
  dX(t) = \mu(X(t))\;dt + \sigma(X(t))\;dW(t)\,,
\end{equation}
$X(t) \in \mathbbm{R}^d$, $\{W(t) \in  \mathbbm{R}^m, ~t\geq0\}$ is
an $m$-dimensional Wiener process, and the functions $\mu\,:\,
\mathbbm{R}^d \rightarrow \mathbbm{R}^d$ and $\sigma\,:\,
\mathbbm{R}^d \rightarrow \mathbbm{R}^{d \times m}$ are sufficiently
smooth. We call $\mu$ the drift coefficient and $\sigma$ the
diffusion coefficient.

Because the Wiener process is non-differentiable, special rules of
stochastic calculus are required when deriving numerical methods for
SDEs.  There are two widely used versions of stochastic calculus,
Ito and Stratonovich~\cite{Kloeden99}.  With Ito calculus, the
solution to SDE \eqref{eq:ito-form} can be represented as an Ito
integral ~\cite{Kloeden99}
\begin{equation}
  \label{eq:integral-form}
  X(t) = X(t_0) + \int_{t_0}^t \mu(X(s))\;ds
  + \int_{t_0}^t \sigma (X(s))\;dW(s), \quad t \in [t_0, T].
\end{equation}
With Stratonovich calculus, the solution to \eqref{eq:ito-form} is
\begin{eqnarray*}
  X(t) &=& X(t_0) + \int_{t_0}^t \underline{\mu}(X(s))\;ds
  + \int_{t_0}^t \sigma (X(s))\;dW(s), \quad t \in [t_0, T], \\
  \underline{\mu}\bigl(X(t)\bigr) &=&  \mu\bigl(X(t)\bigr)
  - \frac{1}{2}\; \sigma\bigl(X(t)\bigr)\;
  \frac{\partial \sigma}{\partial x}\bigl(X(t)\bigr),
\end{eqnarray*}
where $\underline{\mu}$ is the modified drift coefficient.

\subsection{Convergence}
Consider a time discretization of the SDE \eqref{eq:integral-form}
which uses a maximum step size $\delta$ and produces an
approximation $\{Y^{\delta}(t)\}$ of $\{X(t)\}$. The magnitude of
the {\em pathwise} approximation error at a finite terminal time $T$
is measured by the expected absolute value of the difference between
the Ito process and the approximation~\cite{Kloeden99}
\begin{equation*}
  \varepsilon(\delta) = \E\left[\,|X(T)-Y^{\delta}(T)|\,\right]~.
\end{equation*}
The following two definitions of convergence~\cite{Kloeden99} are
useful in the analysis of discretization methods.

\begin{definition} [Strong convergence\cite{Kloeden99}]
A time discrete approximation $Y^{\delta}(t)$ with maximum step size
$\delta$ \emph{converges strongly} to $X$ at time $T$ if
\begin{equation*}
  \lim_{\delta \to 0} \E\left[\,|X(T)-Y^{\delta}(T)|\,\right] = 0,
\end{equation*}
and if there exists a positive constant $C$, which does not depend
on $\delta$, and a finite $\delta_0 > 0$ such that
\begin{equation*}
   \E\left[\,|X(T)-Y^{\delta}(T)|\,\right] \le C \, \delta^{\gamma}
\end{equation*}
for each $\delta \in (0, \delta_0)$, then $Y^{\delta}$ is said to
\emph{converge strongly with order $\gamma > 0$}. \qed
\end{definition}

In many practical situations it is not necessary to have numerical
solutions that accurately approximate each path of an Ito process.
Often one is only interested to accurately compute moments,
probability densities, or other functionals of the Ito process. The
concept of weak convergence~\cite{Kloeden99} describes numerical
accuracy in this situation.

\begin{definition} [Weak convergence\cite{Kloeden99}]
A time discrete approximation $Y^{\delta}(t)$ with maximum step size
$\delta$ \emph{converges weakly} to $X(t)$ at time $T$ as $\delta
\downarrow 0$, with respect to a class $\mathcal{C}$ of polynomials
$g:\mathbb{R}^d \to \mathbb{R}$ if
\begin{equation*}
  \lim_{\delta \to 0} \left|\E\left[g(X(T))\right]
  - \E\left[g(Y^{\delta}(T))\right]\right| = 0,
\end{equation*}
for all $g \in \mathcal{C}$. If there exist a positive constant $C$,
which does not depend on $\delta$, and a finite $\delta_0 > 0$ such
that
\begin{equation*}
   \left|\E\left[g(X(T))\right] - \E\left[g(Y^{\delta}(T))\right] \right|
   \le C\, \delta^{\beta}
\end{equation*}
for each $\delta \in (0, \delta_0)$, then $Y^{\delta}$ is said to
\emph{converge weakly with order $\beta > 0$}. \qed
\end{definition}
\noindent These two convergence criteria lead to the development of
different discretization schemes.

\subsection{Discretization Schemes}
Consider a time discretization $t^0 < t^1 \cdots < t^n < \cdots <
t^N = T$ of the time interval $[t^0, T]$. The stochastic Euler
approximation of the SDE~\eqref{eq:ito-form} is
\begin{equation}
  \label{eq:general-Euler}
  Y_{k}^{n+1} = Y_{k}^{n} + \mu_{k}\;\Delta t^{n} +
  \sum_{j=1}^{m} \sigma_{k,j} \;\Delta W_{j}^{n}~,
  \quad k=1,\cdots,d
\end{equation}
where superscripts denote vector and matrix components. We follow
our convention in writing
\begin{equation*}
\mu_{k} = \mu_{k}(t^{n},Y^{n})~~\mbox{and}~~\sigma_{k,j} =
\sigma_{k,j}(t^{n},Y^{n})~.
\end{equation*}
Here
\begin{equation*}
  \Delta W_{j}^{n} = W_{j}^{t^{n+1}} - W_{j}^{t^{n}}
\end{equation*}
is the $N(0;\,\Delta t^n)$ increment of the $j$th component of the
$m$-dimensional standard Wiener process $W$ on $[t^n,t^{n+1}]$, and
$\displaystyle \Delta W_{j_1}^{n}$ and $\displaystyle \Delta
W_{j_2}^{n}$ are independent for $j_1 \neq j_2$.
 It was shown~\cite{Gikhman72} that the Euler
scheme converges with strong order $\gamma = 0.5$ under Lipschitz
and bounded growth conditions on the coefficients $\mu$ and
$\sigma$.

For weak convergence the random increments $\Delta W^n$ of the
Wiener process can be replaced by other random variables $\Delta
\widehat{W}^n$ which have similar moment properties to the $\Delta
W^n$, but are less expensive to compute~\cite{Kloeden99}. For
instance, in the scalar case $d=m=1$, a weak Euler approximation
with weak order $\beta=1.0$ is
\begin{equation*}
  Y^{n+1} = Y^{n} + \mu\,\Delta t^n + \sigma\,\Delta \widehat{W}^n
\end{equation*}
where $\Delta \widehat{W}^n$ satisfies moment
condition~\cite{Kloeden99}
\begin{equation}
  \label{eq:moment-cond}
  \left|\E\left[\Delta \widehat{W}^n\right]\right|
  + \left|\E\left[(\Delta \widehat{W}^n)^3\right]\right|
  + \left|\E\left[(\Delta \widehat{W}^n)^2\right]
  - \Delta t^n\right| \le C \, (\Delta t^n)^2
\end{equation}
for some constant $C$. A simple example of such a random variable is
the two-point distributed $\displaystyle \Delta \widehat{W}^n$ with
probability
\begin{equation}
  \label{eq:two-point-dist}
  P\left(\Delta \widehat{W}^n = \pm \sqrt{\Delta t^n}\right)
  = \frac{1}{2}~.
\end{equation}

\subsection{The Fully Implicit Euler Scheme}
In the general multi-dimensional case the $k$th component of the
weak Euler scheme has the form
\begin{equation}
  \label{eq:general-weak-Euler}
  Y_{k}^{n+1} = Y_{k}^{n} + \mu_{k}\;\Delta t^{n} +
  \sum_{j=1}^{m} \sigma_{k,j} \;\Delta \widehat{W}_{j}^{n}\,,\quad
  Y_{k}^0=X_0\,,
\end{equation}
where  $\displaystyle \Delta \widehat{W}_{j}^{n}$ satisfies moment
condition~\eqref{eq:moment-cond}. The family of implicit Euler
schemes~\cite{Kloeden99} reads
\begin{equation}
  \label{eq:general-weak-family-implicit-Euler}
  Y_{k}^{n+1} = Y_{k}^{n} + \{\alpha \, \mu_{k}(t^{n+1},Y^{n+1})
  + (1 - \alpha)\,\mu_{k}\} \, \Delta t^n
  + \sum_{j=1}^{m} \sigma_{k,j} \, \Delta \widehat{W}_{j}^{n}~.
\end{equation}
The parameter $\alpha$ here can be interpreted as the degree of
implicitness. With $\alpha=1.0$ it is the implicit Euler scheme,
whereas with $\alpha=0.5$ it represents a stochastic generalization
of the trapezoidal method.

From the definition of Ito stochastic integrals, a meaningful fully
implicit Euler scheme cannot be constructed by making the diffusion
coefficient ($\sigma$) implicit in an equivalent way to the drift
coefficient ($\mu$). To obtain a weakly consistent implicit
approximation it is necessary to appropriately modify the drift
term~\cite{Kloeden99}. Such a family of fully implicit stochastic
Euler schemes is
\begin{align}
  \label{eq:general-weak-family-fully-implicit-Euler}
  Y_{k}^{n+1} = Y_{k}^{n} &+ \left\{\alpha\,
  \overline{\mu}_{k}^{\eta}(t^{n+1},Y^{n+1})
  + (1 - \alpha)\,\overline{\mu}_{k}^{\eta}\right\}
  \Delta t^n  \nonumber \\
  &\quad {} + \sum_{j=1}^{m}\left\{\eta \sigma_{k,j}(t^{n+1},Y^{n+1})
  + (1 - \eta) \sigma_{k,j}\right\} \Delta
  \widehat{W}_{j}^{n}\, ,
\end{align}
where $\displaystyle \Delta \widehat{W}_{j}^{n}$ is as
in~\eqref{eq:two-point-dist} and the corrected drift coefficient
$\displaystyle \overline{\mu}_{k}^{\eta}$ is defined by
\begin{equation}
  \label{eq:corrected-drift}
  \overline{\mu}_{k}^{\eta} = \mu_{k}^{\eta} - \eta \sum_{j=1}^{m}
  \sum_{k=1}^{d} \sigma_{k,j} \frac{\partial \sigma_j}{\partial x_k}\,.
\end{equation}
For $\alpha=\eta=1.0$ the
scheme~\eqref{eq:general-weak-family-fully-implicit-Euler} is the
fully implicit Euler method. For $\eta=0.5$ the corrected drift
$\overline{\mu}_{k}^{\eta} = \underline{\mu}_{k}$ is the corrected
drift of the corresponding Stratonovich equation, and for
$\alpha=0.5$ the
scheme~\eqref{eq:general-weak-family-fully-implicit-Euler} yields
the fully implicit trapezoidal method.

\subsection{The Second Order Weak Taylor Scheme}
In the general multi-dimensional case $d, m = 1, 2, \ldots$ the
$k$th component of the second order weak Taylor scheme
reads~\cite{Kloeden99}
\begin{align}
  \label{eq:general-order2-weakTaylor}
  Y_{k}^{n+1} = Y_{k}^{n} &+ \mu_{k} \, \Delta t^{n} + \frac{1}{2} \,
  L_{0}\,\mu_{k} \, (\Delta t^{n})^2 \nonumber \\
  &\quad {} + \sum_{j=1}^{m}\left\{\sigma_{k,j}
  \, \Delta W_{j}^{n} + L_0 \, \sigma_{k,j} \, I^{(0,j)} + L_j \,
  \mu_k \, I^{(j,0)} \right\}
  + \sum_{j_1,j_2=1}^m L_{j_1} \,
  \sigma_{k,j_2} \, I^{(j_1,j_2)}\,,
\end{align}
where operators $L_0$ and $L_j$ are
\begin{equation*}
  \label{eq:operators}
  L_0 = \frac{\partial}{\partial t} + \sum_{z=1}^{d} \mu_{z} \,
  \frac{\partial}{\partial x_{z}} + \frac{1}{2} \sum_{z,\ell=1}^{d}
  \sum_{h=1}^{m} \sigma_{z,h} \, \sigma_{\ell,h} \,
  \frac{\partial^2}{\partial x_{z} \; \partial x_{\ell}}
  \quad \mbox{and} \quad
  L_j = \sum_{z=1}^{d} \sigma_{z,j} \frac{\partial}{\partial x_{z}}
\end{equation*}
for $j=1,2,\ldots,m$. In addition, the multiple Ito integrals are
abbreviated by
\begin{equation*}
  I^{(j_1,\ldots,j_\ell)} = \int_{t^{n}}^{t^{n+1}} \cdots
  \int_{t^{n}}^{s^2} dW_{j_1}^{s^1} \cdots dW_{j_{\ell}}^{s^{\ell}}.
\end{equation*}

Here we have multiple Ito integrals involving different components
of the Wiener process, which are generally not easy to generate.
Therefore~\eqref{eq:general-order2-weakTaylor} is more of
theoretical interest than of practical use. However, for weak
convergence we can substitute simpler random variables for the
multiple Ito integrals \cite{Kloeden99}. In this way we obtain
from~\eqref{eq:general-order2-weakTaylor} the following simplified
order two weak Taylor scheme with the $k$th component
\begin{align}
  \label{eq:simplified-order2-weakTaylor}
  Y_{k}^{n+1} = Y_{k}^{n} &+ \mu_{k} \, \Delta t^{n} + \frac{1}{2}
  L_{0} \, \mu_{k} \, (\Delta t^{n})^2
  + \sum_{j=1}^{m}\left\{\sigma_{k,j} + \frac{1}{2} \,
  \Delta t^{n} \, (L_{0} \sigma_{k,j} + L_{j} \, \mu_k) \right\} \Delta
  \widehat{W}_j^n \nonumber \\
  & \quad {} + \sum_{j_1,j_2=1}^m L_{j_1} \, \sigma_{k,j_2}
  \left( \Delta \widehat{W}_{j_1}^n \, \Delta
  \widehat{W}_{j_2}^n + V_{j_1,j_2} \right)~.
\end{align}
Here the $\widehat{W}_{j}$ for $j=1,2,\ldots,m$ are independent
random variables satisfying moment conditions
\begin{align}
  \label{eq:moment-cond-2}
  & \left|\E[\Delta \widehat{W}^n]\right|
  + \left|\E\left[(\Delta \widehat{W}^n)^3\right]\right|
  + \left|\E\left[(\Delta \widehat{W}^n)^5\right]\right| \nonumber \\
  & \qquad {} + \left|\E\left[(\Delta \widehat{W}^n)^2\right]
  - \Delta t^n\right|
  + \left|\E\left[(\Delta \widehat{W}^n)^4\right]
  - 3(\Delta t^n)^2 \right|
  \le C \, (\Delta t^n)^3
\end{align}
for some constant $C$. An $N(0;\Delta t^n)$ Gaussian random variable
satisfies the moment condition \eqref{eq:moment-cond-2}, and so does
the three-point distributed $\displaystyle \Delta \widehat{W}^n$
with
\begin{equation}
  \label{eq:three-point-dist}
  P\left(\Delta \widehat{W}^n = \pm \sqrt{3\,\Delta t^n}\right)
  = \frac{1}{6}, \quad P\left(\Delta \widehat{W}^n = 0 \right)
  = \frac{2}{3}.
\end{equation}
The $V_{j_1,j_2}$ are independent two-point distributed random
variables with
\begin{subequations}
\label{eq:V-def}
\begin{equation}
  P\left(V_{j_1,j_2} = \pm \, \Delta t^n \right) = \frac{1}{2}
\end{equation}
for $j_2=1,\ldots,j_{1}-1,$
\begin{equation}
  V_{j_1,j_1} = - \Delta t^n
\end{equation}
and
\begin{equation}
  V_{j_1,j_2} = - V_{j_2,j_1}
\end{equation}
for $j_2=j_{1}+1,\ldots,m$ and $j_1=1,\ldots,m$.
\end{subequations}

\section{Implicit Tau-Leaping-Like Schemes}
\label{sec:newSSA}
We now propose several new fully implicit tau-leaping methods
motivated by the SDE solvers discussed in Section~\ref{sec:SDEs}.

\subsection{The Fully Implicit Tau-Leaping Methods}
We apply the fully implicit weak Euler scheme
\eqref{eq:general-weak-family-fully-implicit-Euler} to the
stochastic chemical kinetic problem. Recall the explicit tau-leaping
method \eqref{eq:exp-tau-leap}. The Poisson variate can be rewritten
as the mean value part plus the variance part of the Poisson
variables. Then the variance term is scaled by the standard
deviation of $a_j(x)$ as below
\begin{equation*}
  \label{eq:scaled-poisson}
  \mathcal{P}_j(a_j(x)\,\tau)=a_j(x)\;\tau + \sqrt{a_j(x)} \;
  \Delta\mathcal{P}_j
\end{equation*}
where the Poisson noise
\begin{equation}
  \label{eq:delta-P}
  \Delta\mathcal{P}_j = \frac{\mathcal{P}_j(a_j(x)\,\tau)
  - a_j(x)\,\tau}{\sqrt{a_j(x)}}
\end{equation}
is close to a normal variable $N(0;\,\tau)$ when $a_j$ is large. The
scheme \eqref{eq:exp-tau-leap} can be written as
\begin{equation}
  \label{eq:exp-tau-scaled}
  X(t+\tau) = x + \sum_{j=1}^M \nu_j \,a_j(x) \,\tau + \sum_{j=1}^M
  \nu_j\,\sqrt{a_j(x)}\,\Delta  \mathcal{P}_j~.
\end{equation}

The weak Euler scheme \eqref{eq:general-weak-Euler}, in vector
notation, reads
\begin{equation}
  \label{eq:general-Euler-vector-notation}
  Y^{n+1} = Y^{n} + \mu \, \Delta t^n + \sum_{j=1}^m \sigma_{j}\,
  \Delta W_{j}^{n}
\end{equation}
where $\sigma_{j}$ is the $j$th column of $\sigma$. We note that
\eqref{eq:exp-tau-scaled} is similar to the Euler
scheme~\eqref{eq:general-Euler-vector-notation} with
\begin{equation}
  \label{eq:coefficients}
  \mu = \sum_{j=1}^M \nu_j\,a_j(x)~,\quad \Delta t^n=\tau~,
  \quad  \sigma_{j}= \nu_j \, \sqrt{a_j(x)}~.
\end{equation}

\subsubsection{The Fully Implicit ``BE--BE'' Method}
The fully implicit ``BE--BE'' tau-leaping method uses the Backward
Euler discretization for both the mean and variance of the Poisson
variables. In~\eqref{eq:general-weak-family-fully-implicit-Euler}
the choice $\alpha=\eta=1$ simplifies the fully implicit weak Euler
scheme to
\begin{equation*}
  \label{eq:fully-weak-implicit-BEBE-first}
  Y^{n+1} = Y^{n} + \overline{\mu}(t^{n+1},Y^{n+1})\,\Delta t^n +
  \sum_{j=1}^{m} \sigma_{j}(t^{n+1},Y^{n+1}) \, \Delta
  \widehat{W}_{j}^{n}
\end{equation*}
where $\displaystyle \Delta \widehat{W}_{j}^{n}$ satisfies moment
condition~\eqref{eq:moment-cond}. Besides the original random
variable $\displaystyle \Delta\widehat{W}_{j}^{n} = \Delta
W_{j}^{n}$, simpler options like~\eqref{eq:two-point-dist} are
possible~\cite{Kloeden99}.

Using~\eqref{eq:coefficients} the corrected drift
coefficient~\eqref{eq:corrected-drift} can be written as
\begin{equation*}
  \overline{\mu} = \mu - \frac{1}{2}\sum_{j=1}^{M}
  \nu_j \left( \sum_{k=1}^{N} \nu_{k,j} \frac{\partial a_j(x)}{\partial
  x_k} \right).
\end{equation*}
Finally the ``BE--BE'' fully implicit tau-leaping method has the
form
\begin{align}
  \label{eq:fully-weak-implicit-BEBE}
  X(t+\tau) = x &+ \tau \sum_{j=1}^M \nu_{j} \,
  \left( a_j\left(X(t+\tau) \right)\right)
  - \frac{\tau}{2} \sum_{j=1}^M \nu_{j} \left( \sum_{k=1}^{N}
  \nu_{k,j} \frac{\partial a_j}{\partial x_k}\left(X(t+\tau)\right)
  \right) \nonumber\\
  & \quad {} + \sum_{j=1}^M \nu_{j} \, \sqrt{a_j\left(X(t+\tau)\right) }
  \, \Delta\widehat{W}_j
\end{align}
where $\displaystyle \Delta \widehat{W}_{j} = \Delta\mathcal{P}_j$.
For large $a_j$, $\displaystyle \Delta\mathcal{P}_j$ is close to a
normal variable and $\displaystyle \Delta \widehat{W}_{j}$ can be
replaced by a random variable with the correct statistics, e.g., as
given by~\eqref{eq:two-point-dist}.

\subsubsection{The Fully Implicit ``TR--TR'' Method}
The fully implicit ``TR--TR'' method uses an implicit trapezoidal
discretization for both the mean of and the variance of the Poisson
variables. The choice $\alpha=\eta=0.5$
in~\eqref{eq:general-weak-family-fully-implicit-Euler} leads to
\begin{equation*}
  Y^{n+1} = Y^{n} + \frac{1}{2} \left \{\overline{\mu}(t^{n+1},Y^{n+1})
  + \overline{\mu} \right \} \, \Delta t^n
  + \frac{1}{2}\sum_{j=1}^{m}\{\sigma_{j}(t^{n+1},Y^{n+1})
  + \sigma_{j}\} \, \Delta \widehat{W}_{j}\, ,
\end{equation*}
where the corrected drift coefficient~\eqref{eq:corrected-drift} is
\begin{equation}
  \label{eq:TRTR-drift}
  \overline{\mu} = \mu - \frac{1}{2}\sum_{j=1}^{m}
  \sum_{k=1}^{d} \sigma_{k,j} \frac{\partial \sigma_j}{\partial x_k}\,,
\end{equation}
and is equivalent to the Stratonovich drift coefficient
$\underline{\mu}$.

From~\eqref{eq:coefficients} the ``TR--TR'' fully implicit
tau-leaping method has the form
\begin{align}
  \label{eq:fully-weak-implicit-TRTR}
  X(t+\tau) = x &+ \frac{\tau}{2} \sum_{j=1}^M \nu_{j}
  \left( a_j\left(X(t+\tau) \right) + a_j(x) \right) \nonumber \\
  & \quad {} - \frac{\tau}{2} \sum_{j=1}^M \nu_{j} \left\{ \frac{1}{4}
  \sum_{k=1}^{N} \nu_{k,j} \left( \frac{\partial a_j(X(t+\tau))}
  {\partial x_k}
  + \frac{\partial a_j(x)}{\partial x_k} \right) \right\} \nonumber \\
  & \qquad {} + \frac{1}{2}\sum_{j=1}^M \nu_{j} \left(
  \sqrt{a_j\left(X(t+\tau)\right) }
  + \sqrt{a_j(x)} \right) \Delta\widehat{W}_j
\end{align}
where the $\displaystyle \Delta \widehat{W}_{j} =
\Delta\mathcal{P}_j$ or, for large $a_j$, can be replaced
by~\eqref{eq:two-point-dist}.

\subsubsection{The Fully Implicit ``BE--TR'' Method}
The fully implicit ``BE--TR'' method uses a backward Euler
discretization for the mean (deterministic) part, and the implicit
trapezoidal discretization for the variance.
In~\eqref{eq:general-weak-family-fully-implicit-Euler} the choice
$\alpha=1.0$ and $\eta=0.5$ simplifies the fully implicit weak Euler
scheme to
\begin{equation*}
  Y^{n+1} = Y^{n} + \overline{\mu}\,(t^{n+1},Y^{n+1})\,\Delta t^n
  + \frac{1}{2}\sum_{j=1}^{m}\{\sigma_{j}(t^{n+1},Y^{n+1})
  + \sigma_{j}(t_{n},Y_{n})\} \, \Delta
  \widehat{W}_{j}\, ,
\end{equation*}
where the corrected drift coefficient~\eqref{eq:corrected-drift} is
equal to~\eqref{eq:TRTR-drift}. From~\eqref{eq:coefficients} the
``BE--TR'' fully implicit tau-leaping method has the form
\begin{align}
  \label{eq:fully-weak-implicit-BETR}
  X(t+\tau) = x &+ \tau \sum_{j=1}^M \nu_{j} \,
  a_j(X(t+\tau)) - \frac{\tau}{4} \sum_{j=1}^M \nu_{j} \left( \sum_{k=1}^{N}
  \nu_{k,j} \frac{\partial a_j(X(t+\tau))}{\partial x_k}
  \right) \nonumber \\
  & \quad {} + \frac{1}{2}\sum_{j=1}^M \nu_{j} \left(
  \sqrt{a_j\left(X(t+\tau)\right) }
  + \sqrt{a_j(x)} \right) \Delta\widehat{W}_j
\end{align}
where the $\displaystyle \Delta \widehat{W}_{j} =
\Delta\mathcal{P}_j$ or, for large $a_j$, can be replaced
by~\eqref{eq:two-point-dist}.

\subsection{Implicit Second Order Weak Taylor Tau-Leaping Methods}
The simplified order two weak Taylor scheme
\eqref{eq:simplified-order2-weakTaylor} motivates the following
family of methods for stochastic kinetic equations:
\begin{align}
  \label{eq:family-implcit-order2-weakTaylor}
  Y_{k}^{n+1} = Y_{k}^{n} &+ \left\{ \alpha \, \mu_{k} (t^{n+1},Y^{n+1})
  + (1-\alpha) \, \mu_{k} \right\} \Delta t^{n} \nonumber \\
  & \quad {} + \frac{1}{2} (1-2\alpha) \left\{ \beta\,L_{0} \, \mu_k(t^{n+1},Y^{n+1})
  + (1-\beta) \, L_0\,\mu_{k} \right\} (\Delta t^{n})^2 \nonumber \\
  & \qquad {} + \frac{1}{2} \sum_{j_1=1,j_2=1}^m L_{j_1} \,
  \sigma_{k,j_2} \left( \Delta \widehat{W}_{j_1}^n \, \Delta
  \widehat{W}_{j_2}^n + V_{j_1,j_2} \right) \nonumber \\
  & \qquad \quad {}+ \sum_{j=1}^{m}\left\{ \sigma_{k,j} +
  \frac{1}{2}(L_0\,\sigma_{k,j} + (1-2\alpha) L_j\,\mu_{k})
  \Delta t^{n} \right\} \Delta \widehat{W}_{j}^{n}~.
\end{align}

\subsubsection{Implicit Second Order Weak SSA with $\alpha=1.0$ and
$\beta=1.0$}
When $\alpha=1.0$ and $\beta=1.0$ the scheme
\eqref{eq:family-implcit-order2-weakTaylor} becomes
\begin{align}
  \label{eq:implcit-order2-weakTaylor-a1b1}
  Y_{k}^{n+1} = Y_{k}^{n} & + \mu_{k}(t^{n+1},Y^{n+1})
  \Delta t^{n} - \frac{1}{2} L_{0}\,\mu_k(t^{n+1},Y^{n+1})
  (\Delta t^{n})^2 \nonumber \\
  & \quad {} + \frac{1}{2} \sum_{j_1=1,j_2=1}^m L_{j_1} \,
  \sigma_{k,j_2} \left( \Delta \widehat{W}_{j_1}^{n} \,
  \Delta \widehat{W}_{j_2}^{n} + V_{j_1,j_2} \right) \nonumber \\
  & \qquad {} + \sum_{j=1}^{m}\left\{ \sigma_{k,j} +
  \frac{1}{2}(L_0\,\sigma_{k,j} - L_j\,\mu_{k})
  \Delta t^{n} \right\} \Delta \widehat{W}_{j}^{n}~.
\end{align}
We apply the implicit order two weak Taylor scheme to the stochastic
chemical kinetic problem in a similar manner to the fully implicit
tau-leaping methods. Note that
\begin{eqnarray}
  \label{eq:operators-all}
  L_0\,\mu &=&  \sum_{k=1}^{d} \mu_{k} \frac{\partial \mu}{\partial x_{k}} +
  \frac{1}{2} \sum_{k,\ell=1}^{d} \sum_{h=1}^{m} \sigma_{k,h} \;
  \sigma_{\ell,h} \frac{\partial^2 \mu}{\partial x_{k} \; \partial
  x_{\ell}}, \quad L_j\,\mu = \sum_{k=1}^{d} \sigma_{k,j} \frac{\partial
  \mu}{\partial x_{k}}, \nonumber \\
  L_0\,\sigma_{j} &=& \sum_{k=1}^{d} \mu_{k} \frac{\partial \sigma_j}{\partial x_{k}}
  + \frac{1}{2} \sum_{k,\ell=1}^{d} \sum_{h=1}^{m} \sigma_{k,h} \;
  \sigma_{\ell,h} \frac{\partial^2 \sigma_j}{\partial x_{k} \; \partial
  x_{\ell}}, \quad \mbox{and} \quad L_{j_1}\sigma_{j_2} = \sum_{k=1}^{d} \sigma_{k,j_1}
  \frac{\partial \sigma_{j_2}}{\partial x_{k}}~.
\end{eqnarray}
From \eqref{eq:coefficients},
\eqref{eq:implcit-order2-weakTaylor-a1b1}, and
\eqref{eq:operators-all} the implicit order two weak tau-leaping SSA
method with $\alpha=1.0$ and $\beta=1.0$ has the form
\begin{align}
  \label{eq:imp-or2weak-a1b1}
  X(t+\tau) = x &+ \tau \sum_{j=1}^M \nu_{j} \,
  \left( a_j\left(X(t+\tau) \right)\right) \nonumber \\
  & {} - \frac{\tau^2}{2} \sum_{j=1}^M \nu_{j}
  \Bigg\{\sum_{k=1}^N \frac{\partial a_j(X(t+\tau))}{\partial x_{k}}
  \left( \sum_{h=1}^{M} \nu_{k,h} a_h(x) \right) \nonumber \\
  & \qquad \qquad \qquad \quad {} + \frac{1}{2}\sum_{k,\ell=1}^N
  \frac{\partial^2 a_j(X(t+\tau))}{\partial x_{k} \,
  \partial x_{\ell}} \left( \sum_{h=1}^{M} \nu_{k,h}
  \nu_{\ell,h} a_h(x) \right) \Bigg\} \nonumber \\
  & {} + \frac{1}{4} \sum_{j_2=1}^{M} \nu_{j_2}
  \frac{1}{\sqrt{a_{j_2}(x)}} \left\{ \sum_{j_1=1}^M
  \sqrt{a_{j_1}(x)} \left( \sum_{k=1}^N \nu_{k,j_1}
  \frac{\partial a_{j_2}(x)}{\partial x_k} \right)
  \left( \Delta\widehat{W}_{j_1} \, \Delta
  \widehat{W}_{j_2} + V_{j_1,j_2} \right) \right\} \nonumber \\
  & {} + \sum_{j=1}^M \left\{ \nu_{j} \sqrt{a_j(x)}
  - \frac{\tau}{2}  \sqrt{a_j(x)}
  \sum_{k=1}^{N} \nu_{k,j} \left( \sum_{h=1}^M
  \nu_{h} \frac{\partial a_h(x)}{\partial x_k} \right)
  \right\} \Delta \widehat{W}_{j} \nonumber\\
  & {}+ \frac{\tau}{4} \sum_{j=1}^{M}
  \frac{\nu_{j}}{4\sqrt{a_j(x)}} \Bigg\{
  \sum_{k=1}^{N} \frac{\partial a_j(x)}{\partial x_k}
  \left( \sum_{h=1}^M \nu_{k,j}\;a_h(x) \right) \nonumber \\
  & \qquad \qquad \qquad \qquad \qquad {}
  - \frac{1}{4 a_j(x)} \sum_{k,\ell=1}^{N}
  \frac{\partial^2 a_j(x)}{\partial x_k \, \partial x_{\ell}}
  \left( \sum_{h=1}^M \nu_{k,h} \; \nu_{\ell,h}\,a_h(x) \right)
  \Bigg\} \Delta \widehat{W}_{j}.
\end{align}

\subsubsection{Implicit Second Order Weak SSA with $\alpha=1.0$ and
$\beta=0.0$}
When $\alpha=1.0$ and $\beta=0.0$ the scheme
\eqref{eq:family-implcit-order2-weakTaylor} reads
\begin{align*}
  Y_{k}^{n+1} = Y_{k}^{n} &+ \mu_{k}(t^{n+1},Y^{n+1})
  \Delta t^{n} - \frac{1}{2} L_{0}\,\mu_k
  (\Delta t^{n})^2 \nonumber \\
  & \quad {} + \frac{1}{2} \sum_{j_1=1,j_2=1}^m L_{j_1} \;
  \sigma_{k,j_2} \left( \Delta \widehat{W}_{j_1}^{n} \Delta \widehat{W}_{j_2}^{n} +
  V_{j_1,j_2} \right) \nonumber \\
  & \qquad {} + \sum_{j=1}^{m}\left\{ \sigma_{k,j} +
  \frac{1}{2}(L_0\,\sigma_{k,j} - L_j\,\mu_{k})
  \Delta t^{n} \right\} \Delta \widehat{W}_{j}^{n}~.
\end{align*}
The corresponding implicit order two weak tau-leaping SSA method has
the form
\begin{align}
  \label{eq:imp-or2weak-a1b0}
  X(t+\tau) & = x + \tau \sum_{j=1}^M \nu_{j} \,
  \left( a_j\left(X(t+\tau) \right)\right) \nonumber \\
  & \quad {} - \frac{\tau^2}{2} \sum_{j=1}^M
  \nu_{j}\left\{\sum_{k=1}^N
  \frac{\partial a_j(x)}{\partial x_{k}}
  \left( \sum_{h=1}^{M} \nu_{k,h} a_h(x) \right)
  + \frac{1}{2}\sum_{k,\ell=1}^N
  \frac{\partial^2 a_j(x)}{\partial x_{k} \,
  \partial x_{\ell}} \left( \sum_{h=1}^{M} \nu_{k,h}
  \nu_{\ell,h} a_h(x) \right)
  \right\} \nonumber \\
  & \quad {} + \frac{1}{4} \sum_{j_2=1}^{M} \nu_{j_2}
  \frac{1}{\sqrt{a_{j_2}(x)}} \left\{ \sum_{j_1=1}^M
  \sqrt{a_{j_1}(x)} \left( \sum_{k=1}^N \nu_{k,j_1}
  \frac{\partial a_{j_2}(x)}{\partial x_k} \right)
  \left( \Delta\widehat{W}_{j_1} \, \Delta
  \widehat{W}_{j_2} + V_{j_1,j_2} \right) \right\} \nonumber \\
  & \quad {}  + \sum_{j=1}^M \left\{ \nu_{j} \sqrt{a_j(x)}
  - \frac{\tau}{2}  \sqrt{a_j(x)}
  \sum_{k=1}^{N} \nu_{k,j} \left( \sum_{h=1}^M
  \nu_{h} \frac{\partial a_h(x)}{\partial x_k} \right)
  \right\} \Delta \widehat{W}_{j} \nonumber\\
  & \quad {}  + \frac{\tau}{4} \sum_{j=1}^{M}
  \frac{\nu_{j}}{4\sqrt{a_j(x)}} \Bigg\{
  \sum_{k=1}^{N} \frac{\partial a_j(x)}{\partial x_k}
  \left( \sum_{h=1}^M \nu_{k,j}\;a_h(x) \right) \nonumber \\
  & \qquad \qquad \qquad \qquad \qquad {}
  - \frac{1}{4 a_j(x)} \sum_{k,\ell=1}^{N}
  \frac{\partial^2 a_j(x)}{\partial x_k \, \partial x_{\ell}}
  \left( \sum_{h=1}^M \nu_{k,h} \; \nu_{\ell,h}\,a_h(x) \right)
  \Bigg\} \Delta \widehat{W}_{j}.
\end{align}

\subsubsection{Implicit Second Order Weak SSA with $\alpha=0.5$}
When $\alpha=0.5$ the scheme
\eqref{eq:family-implcit-order2-weakTaylor} does not depend on
$\beta$. The method reads
\begin{align*}
  Y_{k}^{n+1} = Y_{k}^{n} &+ \frac{1}{2}
  \left\{ \mu_{k}(t^{n+1},Y^{n+1}) + \mu_{k} \right\}
  \Delta t^{n} \nonumber \\
  & \quad {} + \frac{1}{2} \sum_{j_1=1,j_2=1}^m L_{j_1} \;
  \sigma_{k,j_2} \left( \Delta \widehat{W}_{j_1}^{n} \Delta \widehat{W}_{j_2}^{n} +
  V_{j_1,j_2} \right) \nonumber \\
  & \qquad {} + \sum_{j=1}^{m}\left( \sigma_{k,j} +
  \frac{1}{2}L_0\,\sigma_{k,j} \Delta t^{n} \right) \Delta \widehat{W}_{j}^{n}.
\end{align*}
The implicit order two weak tau-leaping SSA method for $\alpha=0.5$
has the form
\begin{align}
  \label{eq:imp-or2weak-a05}
  X(t+\tau) = x &+ \frac{\tau}{2} \sum_{j=1}^M \nu_{j}\,
  \left\{ a_j\left(X(t+\tau)\right) + a_j(x) \right\}
  + \sum_{j=1}^M \nu_{j} \sqrt{a_j(x)} \,
  \Delta \widehat{W}_{j} \nonumber\\
  & + \frac{1}{4} \sum_{j_2=1}^{M} \nu_{j_2}
  \frac{1}{\sqrt{a_{j_2}(x)}} \left\{ \sum_{j_1=1}^M
  \sqrt{a_{j_1}(x)} \left( \sum_{k=1}^N \nu_{k,j_1}
  \frac{\partial a_{j_2}(x)}{\partial x_k} \right)
  \left( \Delta\widehat{W}_{j_1} \, \Delta
  \widehat{W}_{j_2} + V_{j_1,j_2} \right) \right\} \nonumber \\
  & + \frac{\tau}{4} \sum_{j=1}^{M}
  \frac{\nu_{j}}{4\sqrt{a_j(x)}} \Bigg\{
  \sum_{k=1}^{N} \frac{\partial a_j(x)}{\partial x_k}
  \left( \sum_{h=1}^M \nu_{k,j}\;a_h(x) \right) \nonumber \\
  & \qquad \qquad \qquad \qquad \qquad {} - \frac{1}{4 a_j(x)} \sum_{k,\ell=1}^{N}
  \frac{\partial^2 a_j(x)}{\partial x_k \,\partial x_{\ell}}
  \left( \sum_{h=1}^M \nu_{k,h} \; \nu_{\ell,h}\,a_h(x) \right)
  \Bigg\} \Delta \widehat{W}_{j}.
\end{align}

\section{Stability Analysis}\label{sec:analysis}
In this section we perform a theoretical stability analysis of the
fully implicit methods proposed in Section \ref{sec:newSSA}.
Specifically, we take the well established approach \cite{Cao04-2,
Rath05} of applying the methods to the reversible isomerization
model and comparing the discrete results with the available
analytical solution.

\subsection{Reversible Isomerization Model}
Following Rathinam {\it et al.,}~\cite{Cao04-2, Rath05} we consider
the reversible isomerization reaction system
\begin{equation}
  \label{eq:rev-isomer-model}
  S_1 ~ \maprightleft{c_1}{c_2} ~ S_2\,.
\end{equation}
Let $X_t$ denote the population (number of molecules) of $S_1$ at
time $t$, $X^T$ the total population of $S_1$ and $S_2$, and
\begin{equation}
  \lambda=c_1 + c_2\,.
\end{equation}
Usually the case with $c_1=c_2$ is considered. Note that $X^T$ is
constant in time, and therefore the population of $S_2$ at time $t$
is $X^T-X_t$. The deterministic reaction rate equation for this
system is the ODE:
\begin{equation*}
  \frac{dX_t}{dt} = -c_1 X_t + c_2 (X^T - X_t) = -\lambda X_t + c_2
  X^T.
\end{equation*}
Therefore the mean $\E[X_t]$ and variance $\Var[X_t]$ satisfy the
following ODEs:
\begin{align*}
  \frac{d\, \E[X_t]}{dt} & = -\lambda\, \E[X_t] + c_2 X^T, \\
  \frac{d \,\Var[X_t]}{dt} &= -2\lambda\,\Var[X_t] + c_2 X^T + (c_1 -
  c_2)\, \E[X_t].
\end{align*}
As $t$ goes to infinity, the asymptotic value of the exact mean
$\E[X_\infty^*]$ and the exact variance $\Var[X_\infty^*]$
are~\cite{Cao04-2, Hu11}
\begin{equation}
  \label{eq:exact-mean-var}
  \E[X_\infty^*] = \frac{c_2 X^T}{\lambda},~
  \Var[X_\infty^*]=\frac{c_1 c_2 X^T}{\lambda^2}\,.
\end{equation}
%

\subsection{Stability Analysis of the Traditional Tau-leaping Methods}
Recall the explicit tau-leaping method \eqref{eq:exp-tau-leap}.
Applying the explicit tau-leaping method with a fixed step size
$\tau$ to the test problem~\eqref{eq:rev-isomer-model} gives
\begin{equation}
  \label{eq:exp-test}
  X_{n+1} = X_n - \P_1(c_1 \tau X_n) + \P_2(c_2 \tau (X^T-X_n))\,,
\end{equation}
where $X_n$ is the numerical approximation of $X_t$ at time $t_n$.

The following lemma about the conditional probability
from~\cite{Ross06} will prove useful for the derivation.
\begin{lemma}
  \label{lemma:EV}
  If $X$ and $Y$ are random variables, then
  \begin{align*}
    \E[Y] &= \E[\,\E[Y\,|\,X]], \\
    \Var[Y] &= \E[\,\Var[Y\,|\,X]] + \Var[\,\E[Y\,|\,X]].
  \end{align*}
\end{lemma}

By Lemma~\ref{lemma:EV}, the mean of the Eq.~\eqref{eq:exp-test} is
\begin{equation*}
  \E[X_{n+1}] = (1 - \lambda \tau) \,\E[X_n] + c_2 X^T \tau\,.
\end{equation*}
This imposes the stability condition
\begin{equation}
  \label{eq:exp-tau-stability-domain}
  | 1-\lambda \tau| < 1,
\end{equation}
which implies $0 < \lambda \tau < 2$ for the stepsize. For $n
\rightarrow \infty$ we obtain the asymptotic mean
\begin{equation*}
  \E[X_\infty] = \frac{c_2 X^T}{\lambda} = \E[X_\infty^*].
\end{equation*}
For the variance we have
\begin{equation}
  \label{eq:exp-var-eq}
  \Var[X_{n+1}] = (1-\lambda\tau)^2 \,\Var[X_n] + (c_1 - c_2)
  \,\tau \,\E[X_n] + c_2 X^T \tau\,.
\end{equation}
The stable domain for the variance is given by $(1 - \lambda \tau) <
1$ and is the same as~\eqref{eq:exp-tau-stability-domain}. For $n
\rightarrow \infty$ in~\eqref{eq:exp-var-eq}, the asymptotic
variance is
\begin{equation*}
  \Var[X_{\infty}] = \frac{2}{2-\lambda \tau} \, \Var[X_{\infty}^*].
\end{equation*}
Thus the variance given by the explicit tau-leaping method does not
converge to the theoretical value, even if the stability condition
is satisfied. If Eq.~\eqref{eq:exp-tau-stability-domain} is
satisfied, $\Var[X_{\infty}]$ is larger than $\Var[X_{\infty}^*]$.

Similarly, the stability region, asymptotic mean, and asymptotic
variance for the traditional implicit tau-leaping method are
\begin{equation}
  \label{eq:imp-stability-summary}
  \left | \frac{1}{1+\lambda \tau}\right | < 1, \quad
  \E[X_\infty] = \frac{c_2 X^T}{\lambda} = \E[X_\infty^*], \quad
  \Var[X_{\infty}] = \frac{2}{2+\lambda \tau} \, \Var[X_{\infty}^*].
\end{equation}
For the trapezoidal method,
\begin{equation}
  \label{eq:trape-stability-summary}
  \left | \frac{2-\lambda\tau}{2+\lambda\tau} \right | < 1, \quad
  \E[X_\infty] = \frac{c_2 X^T}{\lambda} = \E[X_\infty^*], \quad
  \Var[X_{\infty}] = \frac{c_1 c_2 X^T}{\lambda^2} = \Var[X_{\infty}^*].
\end{equation}
%

\subsection{Stability Analysis of the Fully Implicit Tau-Leaping Methods}
Recall the BE--BE fully implicit
formula~\eqref{eq:fully-weak-implicit-BEBE}
\begin{align*}
  X(t+\tau) = x &+ \sum_{j=1}^M \nu_{j} \Bigg\{
  \tau a_j(X(t+\tau)) - \frac{\tau}{2} \left( \sum_{k=1}^{N}
  \nu_{k,j} \frac{\partial a_j(X(t+\tau))}{\partial x_k} \right)
  \nonumber \\
  & \qquad \qquad \qquad \qquad {}+
  \sqrt{a_j(X(t+\tau))\tau} \left( \frac{\mathcal{P}_j(a_j(x)\,\tau)
  - a_j(x)\,\tau}{\sqrt{a_j(x)}} \right) \Bigg\}.
\end{align*}
We apply the BE--BE tau-leaping methods with a fixed step size
$\tau$ to the test problem~\eqref{eq:rev-isomer-model}. For $N=1$,
$M=2$, $\nu_{1,1}=-1$, $\nu_{1,2}=1$, $a_1(x) = c_1 X$, and $a_2(x)
= c_2(X^T - X)$, we have that
\begin{subequations}
\label{eq:bebe-test-simple}
\begin{align}
  \label{eq:bebe-test-2c}
  X_{n+1} = X_n &- \tau \lambda X_{n+1} + \tau\left( c_2 X^T - \frac{c_1}{2} + \frac{c_2}{2}
  \right)  \\
  \label{eq:bebe-test-2a}
  & \quad {} - \sqrt{X_{n+1}} \left\{ \frac{\P_1(\tau c_1 X_n) - \tau c_1 X_n}{\sqrt{X_n}}
  \right\} \\
  \label{eq:bebe-test-2b}
  & \qquad {} + \sqrt{X^T - X_{n+1}} \left\{ \frac{\P_2(\tau c_2 (X^T - X_n)) - \tau c_2 (X^T - X_n)}{\sqrt{X^T - X_n}} \right\}
\end{align}
\end{subequations}
Derivation of the mean for the simplified
equation~\eqref{eq:bebe-test-simple} is quite intricate due to the
square root in the denominator. In order to derive the stability
region we first employ an inequality condition. Denote by $\E_n
[\,\cdot\,] = \E[\,\cdot | X_n]$; from lemma~\ref{lemma:EV}
$\E[\,\cdot\,] = \E[\E_n[\,\cdot\,]]$. Taking the expectation of
\eqref{eq:bebe-test-2a} leads to
\begin{align*}
&- \E_n \left[ \sqrt{X_{n+1}} \left\{ \frac{\P_1(\tau c_1
X_n) - \tau c_1 X_n}{\sqrt{X_n}} \right\} \right] \\
& \qquad {} \leq \f{1}{2} \, \E_n \left[ X_{n+1} \right] + \f{1}{2} \, \E_n \left[ \f{\left( \P_1(\tau c_1 X_n) - \tau c_1 X_n \right)^2}{X_n}\right]  \\
& \qquad {} = \f{1}{2} \, \E_n \left[ X_{n+1} \right] + \f{1}{2} \f{\Var \left( \P_1 (\tau c_1 X_n)  \right)}{{X_n}} \\
& \qquad {} = \f{1}{2} \, \E_n \left[ X_{n+1} \right] + \f{1}{2}
\tau c_1,
\end{align*}
which implies that
\begin{subequations}
\label{eq:bebe-test-E}
\begin{equation}
\label{eq:bebe-test-2a-E} - \E_n \left[ \sqrt{X_{n+1}} \left\{
\frac{\P_1(\tau c_1 X_n) - \tau c_1 X_n}{\sqrt{X_n}} \right\}
\right] \leq \f{1}{2} \,\E \left[ X_{n+1} \right] + \f{1}{2} \tau
c_1.
\end{equation}
Similarly, the expectation of \eqref{eq:bebe-test-2b} satisfies
\begin{equation}
\label{eq:bebe-test-2b-E} \E \left[ \sqrt{X^T - X_{n+1}} \left\{
\frac{\P_2(\tau c_2 (X^T - X_n)) - \tau c_2 (X^T - X_n)}{\sqrt{X^T -
X_n}} \right\} \right] \leq \f{1}{2} \,\E \left[ X^T- X_{n+1}
\right] + \f{1}{2} \tau c_2.
\end{equation}
\end{subequations}
Plugging \eqref{eq:bebe-test-2a-E} and \eqref{eq:bebe-test-2b-E}
into \eqref{eq:bebe-test-simple} and taking $\E[\,\cdot\,]$ gives
\begin{align*}
\E[X_{n+1}] &\leq \E[X_n] - \tau \lambda \,\E[X_{n+1}] + \tau\left(
c_2 X^T - \frac{c_1}{2} + \frac{c_2}{2} \right) \nonumber \\
& \qquad {} + \f{1}{2} \,\E \left[ X_{n+1} \right] + \f{1}{2} \,\tau
c_1 + \f{1}{2} \,\E \left[ X^T- X_{n+1} \right] + \f{1}{2} \,\tau
c_2,
\end{align*}
which can be simplified to
\begin{equation}
  \label{eq:bebe-stab-eq}
  \E[X_{n+1}] \leq \frac{1}{(1+\lambda \tau)}\E[X_n] + \frac{2\tau c_2
  + 2\tau c_2 X^T + X^T}{(2+2\lambda \tau)}.
\end{equation}
This imposes the sufficient stability condition
\begin{equation}
  \label{eq:bebe-stab-cond}
  \left | \frac{1}{1+\lambda \tau}\right | < 1.
\end{equation}

The second approach for the stability analysis is using the
\emph{Poisson approximation method}.   Recall that the Poisson
random variable can be rewritten as the mean value plus the random
deviation from the mean part
\begin{equation*}
  \mathcal{P}_j(a_j(x)\,\tau)=a_j(x)\tau + \sqrt{a_j(x)} \,
  \Delta\mathcal{P}_j.
\end{equation*}
If $a_j$ is large the Poisson noise $\Delta\mathcal{P}_j$ is close
to a normal variable $N(0;\,\tau)$. In this case the Poisson
variable with mean $a_j(X(t+\tau))\,\tau$ can be approximated by
\begin{equation}
  \label{eq:poisson-approx}
  \P(a_j(X(t+\tau))\,\tau) \approx a_j(X(t+\tau)) \tau +  \sqrt{a_j(X(t+\tau))}\, \Delta \P_j\,.
\end{equation}
With this approximation the ``BE--BE'' fully implicit method has the
alternative form
\begin{equation}
  \label{eq:fully-weak-implicit-BEBE-2}
  X(t+\tau) = x + \sum_{j=1}^M \nu_{j} \P(a_j(X(t+\tau))\,\tau)
  - \frac{\tau}{2} \sum_{j=1}^M \nu_{j} \left( \sum_{k=1}^{N}
  \nu_{k,j} \frac{\partial a_j(X(t+\tau))}{\partial x_k} \right).
\end{equation}
Applying the alternative BE--BE
formula~\eqref{eq:fully-weak-implicit-BEBE-2} with a fixed step size
$\tau$ to the test problem~\eqref{eq:rev-isomer-model} gives
\begin{equation}
  \label{eq:BEBE-test}
  X_{n+1} = X_n - \P_1(c_1 \tau X_{n+1}) + \P_2\left( c_2 \tau
  (X^T - X_{n+1})\right) - \frac{\tau}{2}(c_1 - c_2)\,.
\end{equation}
Denoting by $\E_{n+1}[\,\cdot\,] = \E[\,\cdot\,|X_{n+1}]$ and taking
$\E_{n+1}$ of \eqref{eq:fully-weak-implicit-BEBE-2} leads to
\begin{equation*}
  X_{n+1} = \E_{n+1}[X_n] - c_1 \tau X_{n+1} + c_2 \tau
  (X^T - X_{n+1}) - \frac{\tau}{2}(c_1 - c_2),
\end{equation*}
i.e.,
\begin{equation}
  \label{eq:BEBE-stability-cond-1}
  \E_{n+1}[X_n] =
  (1+\lambda \tau)X_{n+1} - c_2 \tau X^T + \frac{\tau}{2}(c_1 - c_2).
\end{equation}
Then by Lemma ~\ref{lemma:EV} we have
\begin{equation*}
  \label{eq:BEBE-stability-cond-2}
  \E[X_n] = \E[\,\E_{n+1}[X_n]\,]=
  (1+\lambda \tau)\,\E[X_{n+1}] - c_2 \tau X^T + \frac{\tau}{2}(c_1 - c_2).
\end{equation*}
Therefore
\begin{equation}
  \label{eq:BEBE-stability-cond-3}
  \E[X_{n+1}] = \frac{1}{1+\lambda\tau}\,\E[X_n]
  + \frac{\tau}{1+\lambda\tau}\left( c_2 X^T + \frac{c_1 - c_2}{2} \right),
\end{equation}
which imposes the stability condition
\begin{equation}
  \left | \frac{1}{1+\lambda \tau}\right | < 1.
\end{equation}
This approximate stability region is same to the sufficient BE--BE
stability condition~\eqref{eq:bebe-stab-cond} calculated via
inequalities. We conclude that the BE--BE stability is similar to
that of the traditional implicit tau-leaping method for the
reversible isomerization test model.

The Poisson approximation~\eqref{eq:poisson-approx} allows to deduce
the asymptotic mean and variance of the approximate solutions
\eqref{eq:fully-weak-implicit-BEBE-2}. Letting $n \rightarrow
\infty$ in~\eqref{eq:BEBE-stability-cond-3} we obtain
\begin{equation*}
  \E[X_\infty]
  = \frac{1}{\lambda}\left(c_2 X^T + \frac{c_1 - c_2}{2}\right).
\end{equation*}
For $c_1=c_2$ (the common setting of the test problem)
\begin{equation*}
  \E[X_\infty] = \frac{c_2 X^T}{\lambda} = \E[X_\infty^*].
\end{equation*}

The conditional variance of~\eqref{eq:BEBE-test} with respect to
$X_{n+1}$ is
\begin{equation*}
  \Var[X_n|X_{n+1}] = (c_2 - c_1)\tau X_{n+1} - c_2 \tau X^T.
\end{equation*}
Therefore
\begin{equation}
  \label{eq:BEBE-var-1}
  \E[\,\Var[X_n|X_{n+1}]\,] = (c_2 - c_1)\tau \,\E[X_{n+1}] - c_2 \tau X^T.
\end{equation}
The variance of~\eqref{eq:BEBE-stability-cond-1} is
\begin{equation}
  \label{eq:BEBE-var-2}
  \Var[\,\E[X_n|X_{n+1}]\,] = (1+\lambda \tau)^2 \,\Var[X_{n+1}].
\end{equation}
From Lemma~\ref{lemma:EV}, \eqref{eq:BEBE-var-1}, and
\eqref{eq:BEBE-var-2}
\begin{equation*}
  \Var[X_n]
  = (1+\lambda \tau)^2 \,\Var[X_{n+1}]
    + (c_2 - c_1)\tau \,\E[X_{n+1}] - c_2 \tau X^T.
\end{equation*}
Letting $n \rightarrow \infty$
\begin{equation*}
  \Var[X_\infty]
  = (1+\lambda \tau)^2 \,\Var[X_\infty]
  + (c_2 - c_1)\tau \,\E[X_\infty] - c_2 \tau X^T.
\end{equation*}
After replacing the $\displaystyle \E[X_\infty] =
\frac{1}{\lambda}\left(c_2 X^T + \frac{c_1 - c_2}{2}\right)$
\begin{equation*}
  \Var[X_{\infty}]
  = \frac{4c_1 c_2 X^T + (c_1 - c_2)^2}{2 \lambda^2 (2+\lambda \tau)}
\end{equation*}
For $c_1=c_2$ as the $\E[X_\infty]$
\begin{equation*}
  \Var[X_{\infty}]
  = \frac{2c_1 c_2 X^T}{\lambda^2 (2+\lambda \tau)}
  = \frac{2}{2+\lambda \tau} \cdot \frac{c_1 c_2 X^T}{\lambda^2}
  = \frac{2}{2+\lambda \tau} \,\Var[X_{\infty}^*].
\end{equation*}
This asymptotic variance of the approximate
BE--BE~\eqref{eq:fully-weak-implicit-BEBE} is same as that of the
traditional implicit tau-leaping
method~\eqref{eq:imp-stability-summary}.

A similar approach can be used to obtain the stability region, the
asymptotic mean, and the asymptotic variance of the
TR-TR~\eqref{eq:fully-weak-implicit-TRTR} and
BE-TR~\eqref{eq:fully-weak-implicit-BETR} methods. The results are
summarized in Table \ref{tab:implicit-on-isomerization}.

\begin{table}
\centering
\begin{tabular}{|C{2cm}|C{3.5cm}|C{3.5cm}|C{3.5cm}|}
\hline
Method & Stability condition  & $ \E[X_\infty] $ & $\Var[X_{\infty}]$ \\
\hline
BE--BE & $\left | \frac{1}{1+\lambda \tau}\right | < 1$ & $\E[X_\infty^*]$ & $\frac{2}{2+\lambda \tau} \Var[X_{\infty}^*]$ \\
TR--TR & $\left | \frac{2-\lambda\tau}{2+\lambda\tau} \right | < 1$ & $\E[X_\infty^*]$ & $\Var[X_{\infty}^*]$ \\
BE--TR & $\left | \frac{1}{1+\lambda \tau}\right | < 1$ & $\E[X_\infty^*]$ & $ \frac{2}{2+\lambda \tau} \Var[X_{\infty}^*]$ \\
\hline
\end{tabular}
\caption{\label{tab:implicit-on-isomerization}Behavior of fully
implicit methods applied to the reversible isomerization problem.
All methods are unconditionally stable and yield the exact
asymptotic mean. TR--TR provides the exact asymptotic variance as
well.}
\end{table}
%

\subsection{Stability Analysis of the Implicit Second Order Tau-Leaping Methods}
Application of the implicit second order method with $\alpha = 1.0$
and $\beta = 1.0$ \eqref{eq:imp-or2weak-a1b1} to the test
problem~\eqref{eq:rev-isomer-model} yields
\begin{equation}
\label{ex:2nd-imp-a1b1-test} X_{n+1} = X_n + \tau (c_2 X^T - \lambda
X_{n+1}) + \f{1}{4} \left( r_1 - r_2 - r_3 + r_4 \right) + r_5 + r_6
+ \f{\lambda \tau^2}{2} (c_2 X^T - \lambda X_n),
\end{equation}
with
\begin{align*}
r_1 &= \f{\left\{\P_1(\tau c_1 X_n) - \tau c_1 X_n\right\}^2}{X_n} + c_1 V_{1,1}, \nonumber \\
r_2 &= \f{\left\{\P_2(\tau c_2(X^T - X_n)) - \tau c_2(X^T-X_n)\right\}^2}{X^T-X_n} + c_2 V_{2,2}, \nonumber \\
r_3 &= \frac{\left\{\P_1(\tau c_1 X_n) - \tau c_1 X_n\right\}\cdot\left\{\P_2(\tau c_2 (X^T-X_n)) - \tau c_2 (X^T-X_n)\right\}}{X_n} + \sqrt{\f{c_1 c_2(X^T- X_n)}{X_n} } V_{2,1}, \nonumber \\
r_4 &= \frac{\left\{\P_1(\tau c_1 X_n) - \tau c_1 X_n\right\}\cdot\left\{\P_2(\tau c_2 (X^T-X_n)) - \tau c_2 (X^T-X_n)\right\}}{X^T - X_n} + \sqrt{\f{c_1 c_2 X_n}{X^T-X_n}} V_{1,2}, \nonumber \\
r_5 &= \left( 1+ \f{\lambda \tau}{2} \right) \left\{ \P_2(\tau c_2(X^T -X_n)) - \tau c_2(X^T-X_n) - \P_1(\tau c_1X_n) + \tau c_1 X_n \right\}, \nonumber \\
r_6 &= \f{\tau}{16} \Bigg[( \lambda X_n - c_2 X^T) \bigg\{
\f{\P_1(\tau c_1 X_n) - \tau c_1 X_n}{X_n} + \f{\P_2 (\tau
c_2(X^T-X_n)) - \tau c_2(X^T-X_n)}{X^T-X_n} \bigg\} \Bigg].
\nonumber
\end{align*}
where The $V_{j_1,j_2}$ are independent two-point distributed random
variables as \eqref{eq:V-def}. In order to derive the mean of
equation
\eqref{ex:2nd-imp-a1b1-test}, we first compute $\E_n[r_1],...,\E_n[r_6]$.   Using $\E_n[V_{1,1}] = - \tau$, 
\begin{align*}
\E_n[r_1] = \E_n \left[ \f{\left\{\P_1(\tau c_1 X_n) - \tau c_1
X_n\right\}^2}{X_n} + c_1 V_{1,1} \right] =\frac{\Var\left(
\P_1(\tau c_1 X_n)\right)}{X_n} - \tau c_1 =0\,.
\end{align*}
Similarly, $\E_n[r_j] = 0$ for $j = 2, \ldots, 6$. Therefore
\begin{equation*}
  (1+ \lambda \tau ) \E_n[X_{n+1}] = \left(  1 - \f{\lambda^2
  \tau^2}{2}\right) \E_n[X_n] + \tau c_2 X^T \left( 1+ \f{\lambda
  \tau}{2}\right).
\end{equation*}
From Lemma~\ref{lemma:EV}, the mean of the numerical solution
satisfies
\begin{equation}
  \label{eq:imp-or2weak-a1b1-stab-eq}
  \E[X_{n+1}] = \left( \f{2 - \lambda^2 \tau^2}{2+2\lambda\tau}\right)
  \E[X_n] + \frac{\tau c_2 X^T (2 + \lambda \tau)}{2+2\lambda\tau},
\end{equation}
which implies the stability restriction
\begin{equation}
  \label{eq:imp-or2weak-a1b1-stab-cond}
  \left| \f{2 - \lambda^2 \tau^2}{2 + 2 \lambda \tau }\right| < 1
  \quad \Rightarrow \quad 0<\lambda \tau<1+\sqrt{5}\,.
\end{equation}
The second order weak Taylor method with $\alpha=1.0$ and
$\beta=1.0$ is {\em conditionally stable}. For the asymptotic mean
of the second order weak Taylor method with $\alpha=1.0$ and
$\beta=1.0$, let $n \to \infty$ in
\eqref{eq:imp-or2weak-a1b1-stab-eq}. Then we obtain
\begin{equation}
  \label{eq:imp-or2weak-a1b1-asymp-mean}
  \E[X_\infty] = \f{c_2 X^T}{\lambda} = \E[X^*_\infty],
\end{equation}
which is equal to its exact value \eqref{eq:exact-mean-var}.

The stability condition and the asymptotic mean for the implicit
second order with $\alpha=1.0$ and $\beta=0.0$
\eqref{eq:imp-or2weak-a1b0} are calculated in a similar manner, and
the results are the same as \eqref{eq:imp-or2weak-a1b1-stab-cond}
and \eqref{eq:imp-or2weak-a1b1-asymp-mean}.

Application of the implicit second order method with $\alpha = 0.5$
\eqref{eq:imp-or2weak-a05} to the test
problem~\eqref{eq:rev-isomer-model} gives
\begin{equation}
  \label{ex:2nd-imp-a05-test}
  X_{n+1} = X_n + \frac{\tau}{2} (2c_2 X^T - \lambda X_{n+1} - \lambda X_n)
  + \f{1}{4} \left( r_1 - r_2 - r_3 + r_4 \right) + r_5 + r_6,
\end{equation}
with
\begin{align*}
r_1 &= \f{\left\{\P_1(\tau c_1 X_n) - \tau c_1 X_n\right\}^2}{X_n} + c_1 V_{1,1}, \nonumber \\
r_2 &= \f{\left\{\P_2(\tau c_2(X^T - X_n)) - \tau c_2(X^T-X_n)\right\}^2}{X^T-X_n} + c_2 V_{2,2}, \nonumber \\
r_3 &= \frac{\left\{\P_1(\tau c_1 X_n) - \tau c_1 X_n\right\}\cdot\left\{\P_2(\tau c_2 (X^T-X_n)) - \tau c_2 (X^T-X_n)\right\}}{X_n} + \sqrt{\f{c_1 c_2(X^T- X_n)}{X_n} } V_{2,1}, \nonumber \\
r_4 &= \frac{\left\{\P_1(\tau c_1 X_n) - \tau c_1 X_n\right\}\cdot\left\{\P_2(\tau c_2 (X^T-X_n)) - \tau c_2 (X^T-X_n)\right\}}{X^T - X_n} + \sqrt{\f{c_1 c_2 X_n}{X^T-X_n}} V_{1,2}, \nonumber \\
r_5 &= \P_2(\tau c_2(X^T -X_n)) - \tau c_2(X^T-X_n) - \P_1(\tau c_1X_n) + \tau c_1 X_n, \nonumber \\
r_6 &= \f{\tau}{16} \Bigg[( \lambda X_n - c_2 X^T) \bigg\{
\f{\P_1(\tau c_1 X_n) - \tau c_1 X_n}{X_n} + \f{\P_2 (\tau
c_2(X^T-X_n)) - \tau c_2(X^T-X_n)}{X^T-X_n} \bigg\} \Bigg].
\nonumber
\end{align*}
Similar to the calculation for the implicit second order weak SSA
with $\alpha=1.0$ and $\beta=1.0$, taking expected value $\E_n$ and
then $\E$ gives
\begin{equation}
  \label{eq:imp-or2weak-a05-stab-eq}
  \E[X_{n+1}] = \left( \f{2 - \lambda\tau}{2 + \lambda\tau}\right)
  \E[X_n] + \frac{2 \tau c_2 X^T}{2+\lambda\tau}.
\end{equation}
The asymptotic stability of $\E[X_n]$ requires
\begin{equation}
  \label{eq:imp-or2weak-a05-stab-cond}
  \left|  \f{2 - \lambda \tau }{ 2 + \lambda \tau} \right| < 1
  \quad \Rightarrow \quad 0 < \lambda \tau\,.
\end{equation}
Because $\lambda \tau$ is always greater than zero, the second order
weak Taylor methods with $\alpha=0.5$ is {\em unconditionally
stable}. The condition \eqref{eq:imp-or2weak-a05-stab-cond} is the
same as that \eqref{eq:trape-stability-summary} of the trapezoidal
tau-leaping method. Letting $n \to \infty$ we have
\begin{equation*}
  \E[X_\infty] = \f{c_2 X^T}{\lambda} = \E[X^*_\infty],
\end{equation*}
which is equal to its exact value \eqref{eq:exact-mean-var}.

Deriving analytically the asymptotic variances for the second order
weak Taylor methods becomes a very intricate task. For the variance
of the implicit second order method with $\alpha = 0.5$
\eqref{eq:imp-or2weak-a05} to the test
problem~\eqref{eq:rev-isomer-model}, we still use the fact
\begin{equation*}
  \Var[X_{n+1}] = \E[\,\Var[X_{n+1} | X_n]\,]
  + \Var [\,\E[X_{n+1} | X_n]\,]
\end{equation*}
using Lemma \eqref{lemma:EV}. By \eqref{eq:imp-or2weak-a05-stab-eq},
\begin{equation*}
  \Var[\,\E[X_{n+1} | X_n]\,] = \left( \f{2 - \lambda \tau}{2+
  \lambda \tau}\right)^2 \Var[X_n]
\end{equation*}
To calculate the term $\E[\,\Var[X_{n+1} | X_n]\,]$, we should
consider the expectation of the variance of
\eqref{ex:2nd-imp-a05-test}. This involves the estimation of
$\E[\f{1}{X_n}]$ and $\E[\f{1}{X^T - X_n}]$ which cannot be obtained
simply. This intractable calculation will be analyzed in future
work.

\section{Experimental Results} \label{sec:results}
This section presents numerical results for the new implicit
tau-leaping methods applied to three different systems. A fixed
stepsize strategy  is used in each simulation for all methods; this
allows for a clean comparison of the performance of different
algorithms.

\subsection{The Decaying-Dimerizing Reaction Set}
\begin{figure}[!t]
  \begin{center}
  \center \scalebox{1.0}
  {\includegraphics{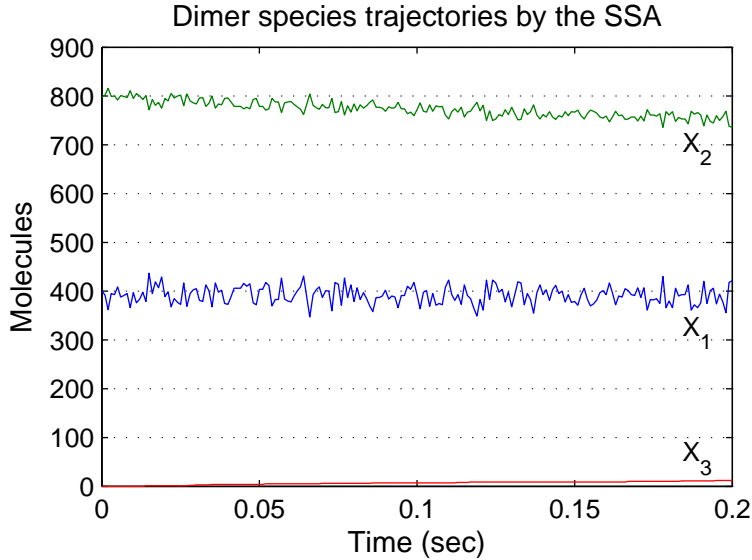}}
  \caption{Time evolution of the numbers of molecules in the
  decaying-dimerizing problem \eqref{eqn:decaying-dimerizing}.
  The simulation is carried out using Gillespie's SSA method.}
  \label{fig:dimer_traj_ssa}
  \end{center}
\end{figure}

The decaying-dimerizing system \cite{Rath05} consists of three
species $S_1$, $S_2$, and $S_3$ and four reactions
\begin{equation}
 \label{eqn:decaying-dimerizing}
  \begin{split}
    &S_1 \; \mapright{c_1}\; 0, \\
    &S_1+S_1 \; \maprightleft{c_2}{c_3} \; S_2, \\
    &S_2 \; \mapright{c_4} \; S_3.
    \end{split}
\end{equation}
We chose the following values for the parameters
\begin{equation*}
  c_1=1, \quad c_2=10, \quad c_3=1000, \quad c_4=0.1,
\end{equation*}
which will render the problem stiff. The propensity functions are
\begin{equation*}
  a_1=X_1,\; a_2=5X_1(X_1-1),\; a_3=1000X_2,\; a_4=0.1X_2,
\end{equation*}
where $X_i$ denotes the number of molecules of species $S_i$. The
initial conditions are
\begin{equation*}
  X_1(0)=400, \quad X_2(0)=798, \quad X_3(0)=0~\textrm{[molecules]}.
\end{equation*}
The final time is $T=0.2$ seconds. Figure~\ref{fig:dimer_traj_ssa}
shows the species evolution for the reaction
set~\eqref{eqn:decaying-dimerizing} solved with the original SSA.

\begin{figure}[!t]
  \centering
  \subfigure[ Histograms obtained with Gillespie's SSA,
  and with the traditional explicit, implicit, and trapezoidal tau-leaping methods.]{
    \includegraphics[width=3.1in]{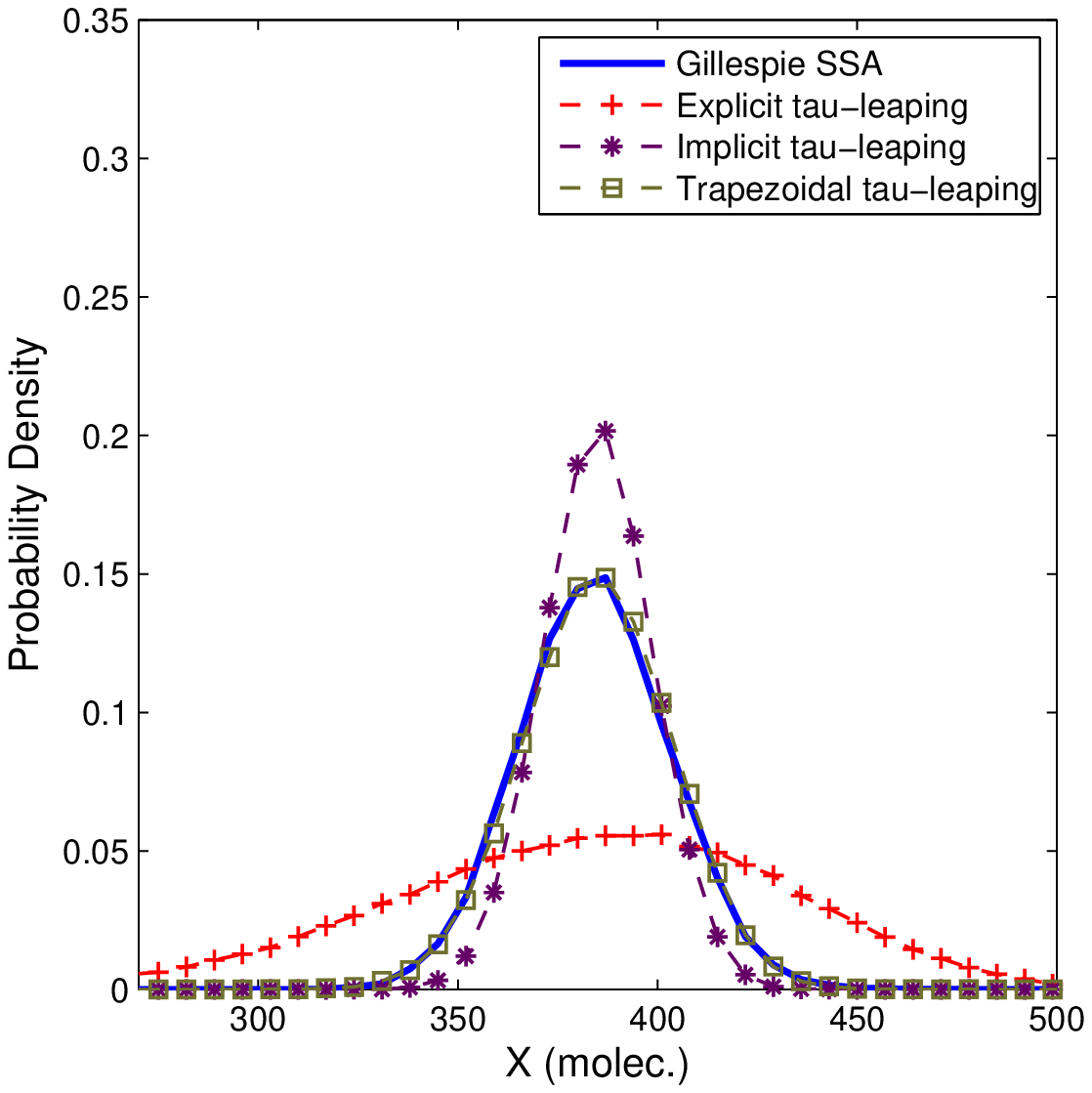}
  }
  \subfigure[ Histograms obtained with Gillespie's SSA,
  the new fully implicit methods, and the new implicit order two weak
  Taylor tau-leaping methods.]{
    \includegraphics[width=3.1in]{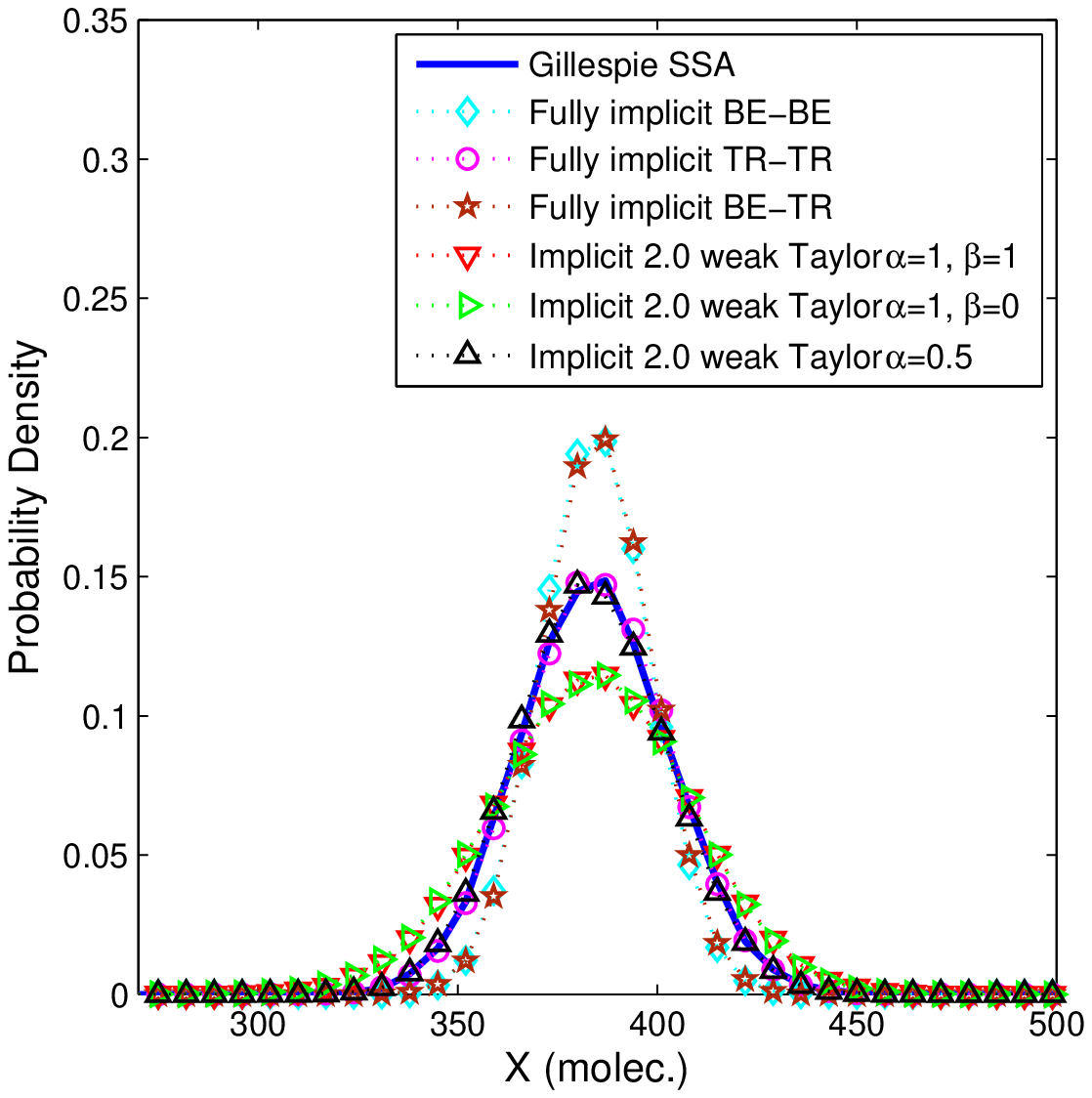}
  }
  \caption{The histograms  of the number of molecules $X_1$ at the final time for
  the decaying-dimerizing reaction system  \eqref{eqn:decaying-dimerizing}.
  All histograms are based on 100,000 runs of the corresponding methods with a fixed stepsize $\tau=2
  \times 10^{-4}$ seconds.}
  \label{fig:dimer_hist}
\end{figure}

In order to compare the solutions given by different methods we
consider histograms of $X_1$, the number of molecules of $S_1$, at
the final time $T=0.2$ seconds. Specifically, an ensemble of
simulation results is carried out for each method, and the final
distribution of the numerical $X_1$ is plotted as a histogram from
100,000 independent simulations.

Figure~\ref{fig:dimer_hist}(a) shows the histograms of $X_1$ for the
decaying-dimerizing system \eqref{eqn:decaying-dimerizing} simulated
with Gillespie's SSA and with the traditional explicit tau-leaping,
implicit tau-leaping, and trapezoidal tau-leaping methods. A fixed
stepsize $\tau=2 \times 10^{-4}$ seconds is used.
Figure~\ref{fig:dimer_hist}(b) also shows the histograms generated
with Gillespie's SSA, and with the methods proposed herein: fully
implicit BE--BE, TR--TR, BE--TR, implicit order two weak Taylor with
$\alpha=1.0$ and $\beta=1.0$, $\alpha=1.0$ and $\beta=0.0$, and
$\alpha=0.5$. The same fixed stepsize $\tau=2 \times 10^{-4}$ is
used.

Figures~\ref{fig:dimer_hist} (a) and (b) reveal that the histograms
of the trapezoidal tau-leaping method, fully implicit TR--TR method,
and implicit order two weak Taylor method with $\alpha=0.5$ are
closer to the reference (SSA) histogram than those of other methods,
for the specific time step chosen.

The explicit method gives very unstable and varying results. Other
implicit order two weak Taylor methods with $\alpha=1.0$ provoke a
little wide varying results, but those escape the damping effect
such as implicit tau-leaping method in Figure~\ref{fig:dimer_hist}
(a). From the stability analysis, we have proved that the implicit
order two weak Taylor methods with $\alpha=1.0$ are unstable for
large stepsizes, and these experimental results confirm the
conditional stability.

In order to numerically assess the accuracy of each method, we carry
out simulations with different stepsizes, and obtain the
corresponding histograms. For each method and step size the
numerical errors are quantified by the difference between the
numerical histograms and the reference (SSA) histogram. Two metrics
of the difference are employed: the  Kullback-Leibler (K-L)
divergence~\cite{Emmert08} and the distance metric.

The K-L divergence is a non-commutative measure of the difference
between two probability distributions $P$ and $Q$, typically $P$
representing the ``true'' distribution and $Q$ representing
arbitrary probability distribution. Therefore we set $P$ to be the
distribution obtained from SSA, and $Q$ the distribution obtained
with one of the other formulae. The K-L divergence is defined to be
\begin{equation}
D_{KL}(P||Q)=\sum_{i}P(i)\log{P(i)\over Q(i)}
\end{equation}
where $Q(i)\neq0$, and the summation is taken over the histogram
bins. Smaller values of K-L divergence represent more similar
distributions. Because K-L divergence is not useful when there
exists zeros for $Q$, we also use the distance metric, which
measures the difference between two distributions by
\begin{equation}
Dist=\sum_{i}\Delta X \cdot |P(i)-Q(i)|\,. \label{eq:dist}
\end{equation}
Here $\Delta X$ is the bin size of the histogram.

\begin{table*}[h!]
\caption{The mean, variance, K-L divergence, and distance for $X_1$
at $T=0.2$ based on 100,000 samples for different stepsizes of the
decaying-dimerizing reaction
system~\eqref{eqn:decaying-dimerizing}.}
\vskip 0.10in
\centering
\linespread{1.00}
\footnotesize
\begin{tabular}{|C{3cm}|C{2.5cm}|C{1.8cm}|C{1.8cm}|C{1.8cm}| C{1.8cm} |} \hline
\multicolumn{2}{|c|}{} & \multicolumn{4}{|c|}{\textbf{Stepsize
($\tau$ in seconds)}} \\ \hline
\multicolumn{1}{|c|}{\textbf{Method}} &
\multicolumn{1}{|c|}{\textbf{Metrics}} &
  \multicolumn{1}{|c|}{$8 \times 10^{-4}$} &
  \multicolumn{1}{|c|}{$4 \times 10^{-4}$} &
  \multicolumn{1}{|c|}{$2 \times 10^{-4}$} &
  \multicolumn{1}{|c|}{$1 \times 10^{-4}$} \\ \hline \hline
\multicolumn{1}{|c|}{Gillespie} & \multicolumn{1}{|c|}{Mean} & \multicolumn{4}{|c|}{387.19} \\
\multicolumn{1}{|c|}{SSA} & \multicolumn{1}{|c|}{Variance} &
\multicolumn{4}{|c|}{349.87} \\ \hline
Explicit       & Mean           & $\infty$ & $\infty$ & 384.71  & 386.92 \\
tau-leaping    & Variance       & $\infty$ & $\infty$ & 2503.30 & 614.64 \\
               & K-L div.       & $\infty$ & $\infty$ & 0.740   & 0.092  \\
               & Distance       & $\infty$ & $\infty$ & 8.799   & 2.665  \\ \hline
Implicit       & Mean           & 387.95   & 387.86   & 387.92  & 387.81 \\
tau-leaping    & Variance       & 79.42    & 128.46   & 185.93  & 242.84 \\
               & K-L div.       & 0.329    & 0.176    & 0.080   & 0.030  \\
               & Distance       & 6.689    & 4.829    & 3.156   & 1.817  \\ \hline
Trapezoidal    & Mean           & 387.63   & 387.70   & 387.73  & 387.60 \\
tau-leaping    & Variance       & 351.29   & 346.61   & 346.38  & 347.24 \\
               & K-L div.       & 0.004    & 0.004    & 0.002   & 0.002  \\
               & Distance       & 0.617    & 0.584    & 0.444   & 0.370  \\ \hline \hline
Fully implicit & Mean           & 387.27   & 387.35   & 387.37  & 387.49 \\
BE--BE         & Variance       & 79.02    & 128.21   & 184.31  & 239.5  \\
               & K-L div.       & 0.329    & 0.174    & 0.080   & 0.031  \\
               & Distance       & 6.583    & 4.744    & 3.078   & 1.859  \\ \hline
Fully implicit & Mean           & 387.26   & 387.43   & 387.51  & 387.61 \\
TR--TR         & Variance       & 348.09   & 343.71   & 344.10  & 346.91 \\
               & K-L div.       & 0.003    & 0.002    & 0.001   & 0.001  \\
               & Distance       & 0.413    & 0.312    & 0.296   & 0.276  \\ \hline
Fully implicit & Mean           & 387.63   & 387.63   & 387.77  & 387.59 \\
BE--TR         & Variance       & 79.54    & 127.60   & 187.74  & 241.69 \\
               & K-L div.       & 0.326    & 0.177    & 0.077   & 0.030  \\
               & Distance       & 6.604    & 4.818    & 3.031   & 1.905  \\ \hline\hline
Implicit 2.0   & Mean           & $\infty$ & $\infty$ & 386.49  & 387.12 \\
weak Taylor    & Variance       & $\infty$ & $\infty$ & 584.70  & 407.24 \\
($\alpha=1,\beta=1$) & K-L div. & $\infty$ & $\infty$ & 0.076   & 0.007  \\
               & Distance       & $\infty$ & $\infty$ & 2.426   & 0.672  \\ \hline
Implicit 2.0   & Mean           & $\infty$ & $\infty$ & 386.07  & 387.03 \\
weak Taylor    & Variance       & $\infty$ & $\infty$ & 591.80  & 409.78 \\
($\alpha=1,\beta=0$) & K-L div. & $\infty$ & $\infty$ & 0.080   & 0.007  \\
               & Distance       & $\infty$ & $\infty$ & 2.455   & 0.726  \\ \hline
Implicit 2.0   & Mean           & 387.29   & 387.26   & 386.44  & 386.25 \\
weak Taylor    & Variance       & 356.93   & 350.17   & 348.72  & 348.89 \\
($\alpha=0.5$) & K-L div.       & 0.004    & 0.003    & 0.002   & 0.002  \\
               & Distance       & 0.625    & 0.421    & 0.386   & 0.318  \\ \hline
\end{tabular}
\label{table:dimer_compare}
\end{table*}

Table~\ref{table:dimer_compare} shows these metrics based on 100,000
samples generated by different methods for fixed stepsizes
$\tau=(8/k) \times 10^{-4}$ where $k=1,2,4,8$. The results show that
the mean is accurately computed by all accelerated methods. However,
the variance and distance are different for each formula. For
example, the explicit tau formula becomes very unstable for a
stepsize of $4 \times 10^{-4}$ seconds. The implicit tau-leaping,
BE--BE, BE--TR are far superior to explicit tau, but those formulae
produce smaller variances compared to the variance of the exact SSA
that is called as damping effect.

Three methods (the trapezoidal-tau, the fully implicit TR--TR, and
the implicit second order weak Taylor with $\alpha=0.5$) generate
accurate variance results even with large stepsizes. The fully
implicit TR--TR results are the most accurate among all methods for
similar time steps, as demonstrated by the smaller distance to the
reference histogram in Table~\ref{table:dimer_compare}. The implicit
second order weak Taylor methods with $\alpha=1.0$ are accurate
until they become unstable for large stepsizes.

\begin{table*}[t]
\caption{Elapsed CPU times (in seconds) for each method and time
step for 100,000 simulations of the decaying-dimerizing reaction
system  \eqref{eqn:decaying-dimerizing}.}
\vskip 0.10in
\centering
\linespread{1.00}
\footnotesize
\begin{tabular}{|C{6.2cm}|C{2cm}|C{2cm}|C{2cm}| C{2cm} |} \hline
\multicolumn{1}{|c|}{\textbf{CPU time (seconds)}} &
\multicolumn{4}{|c|}{\textbf{Stepsize ($\tau$ in seconds)}} \\
\hline \multicolumn{1}{|c|}{\textbf{Method}} &
  \multicolumn{1}{|c|}{$8 \times 10^{-4}$} &
  \multicolumn{1}{|c|}{$4 \times 10^{-4}$} &
  \multicolumn{1}{|c|}{$2 \times 10^{-4}$} &
  \multicolumn{1}{|c|}{$1 \times 10^{-4}$} \\ \hline \hline
\multicolumn{1}{|c|}{Gillespie SSA} & \multicolumn{4}{|c|}{16210.13}
\\ \hline Explicit tau-leaping                            & 27.32 &
46.91 & 130.55 & 260.24  \\ \hline Implicit tau-leaping
& 170.57 & 340.58 & 657.51 & 1389.29  \\ \hline Trapezoidal
tau-leaping                         & 180.42 & 350.66 & 688.98 &
1301.21 \\ \hline Fully implicit BE--BE                           &
344.98 & 686.49 & 1395.1 & 2638.74 \\ \hline Fully implicit TR--TR
& 377.06 & 746.24 & 1400.96 & 2752.39 \\ \hline Fully implicit
BE--TR                           & 340.65 & 690.56 & 1373.31 &
2657.25 \\ \hline Implicit 2.0 weak Taylor ($\alpha=1,\beta=1$) &
398.23 & 784.43 & 1587.69 & 3121.32 \\ \hline Implicit 2.0 weak
Taylor ($\alpha=1,\beta=0$) & 391.31 & 765.39 & 1532.98 & 3076.23 \\
\hline Implicit 2.0 weak Taylor ($\alpha=0.5$)       & 381.34 &
752.84 & 1425.83 & 2798.54 \\ \hline
\end{tabular}
\label{table:dimer_cpu}
\end{table*}

\begin{figure}[h!]
  \begin{center}
  \center
  \scalebox{0.7}{\includegraphics{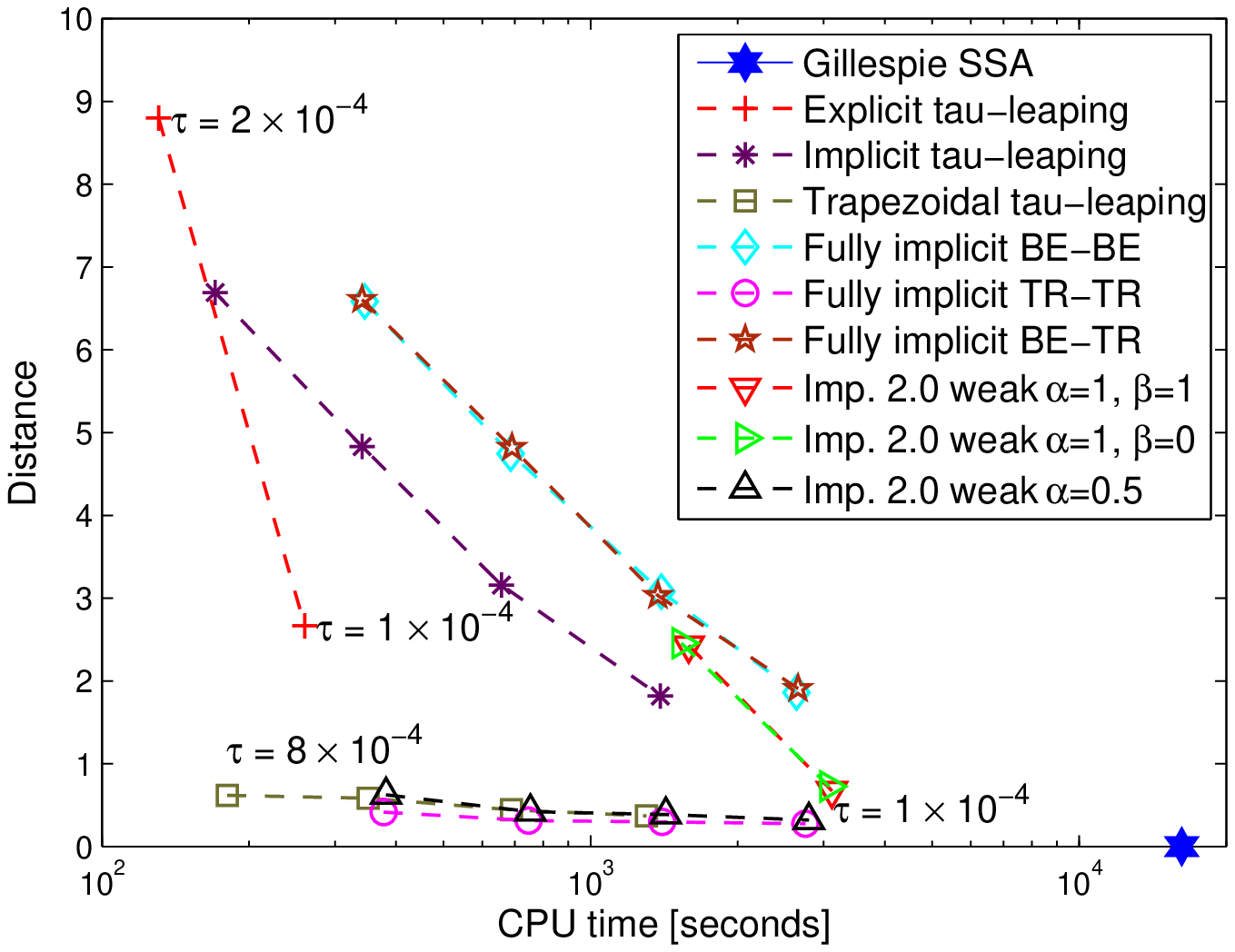}}
  \caption{Relationship between solution accuracy (measured by the distance
  \eqref{eq:dist} between the accelerated method and the SSA produced histograms)
  and CPU time for different methods applied to the decaying-dimerizing
  reaction system~\eqref{eqn:decaying-dimerizing}.}
  \label{fig:dimer_dist_cpu}
  \end{center}
\end{figure}

The elapsed CPU times for each method are presented in
Table~\ref{table:dimer_cpu}. Figure~\ref{fig:dimer_dist_cpu}
considers the relationship between accuracy and computation time for
each of the accelerated methods. From the figure, the trapezoidal
tau-leaping, the fully implicit TR--TR, and the implicit second
order weak Taylor with $\alpha=0.5$ methods generate accurate
solutions with a large step size ($\tau = 8 \times 10^{-4}$ seconds)
and in a short CPU time. For comparison, 100,000 simulations using
the SSA took 16,210 CPU seconds, while 100,000 simulations of the
fully implicit TR--TR took only 377 seconds (2.3\% of the SSA time)
and provided an accurate solution (distance value is only 0.276).
The implicit second order weak Taylor method of the $\alpha=0.5$
with $\tau = 8 \times 10^{-4}$ fixed step took 381 seconds and
produced results of similar accuracy.

\subsection{Schl\"{o}gl Reaction Set}

The Schl\"{o}gl reaction model~\cite{Cao04-2} is a simple but famous
bistable system. The system contains four reactions
\begin{equation}
 \label{eqn:schlogl_system}
  \begin{split}
    &B_1+2S \; \maprightleft{c_1}{c_2} \; 3S, \\
    &B_2 \; \maprightleft{c_1}{c_2} \; S,
    \end{split}
\end{equation}
where $B_1$ and $B_2$ are buffered species whose populations are
assumed to remain constant over the time interval.
\begin{equation*}
  c_1=3\times10^{-7}, \quad c_2=10^{-4}, \quad c_3=10^{-3}, \quad
  c_4=3.5, N_1 = 1\times10^5, \quad N_2 = 2\times10^5.
\end{equation*}
which will render the bistable system. Hence the propensity
functions are given by
\begin{equation*}
  a_1=\frac{c_1}{2}N_1X(X-1), \quad a_2=\frac{c_2}{6}X(X-1)(X-2), \quad
  a_3=c_3 N_2, \quad a_4=c_4 X
\end{equation*}
where $X$ denotes the number of molecules of species $S$. Initial
condition $X(0) = 250$ at $T=0$, and final time $T=4$ second.

\begin{figure}[!t]
  \centering
  \subfigure[Histograms obtained with Gillespie's SSA,
  and with the traditional explicit, implicit, and trapezoidal tau-leaping methods.]{
    \includegraphics[width=3.1in]{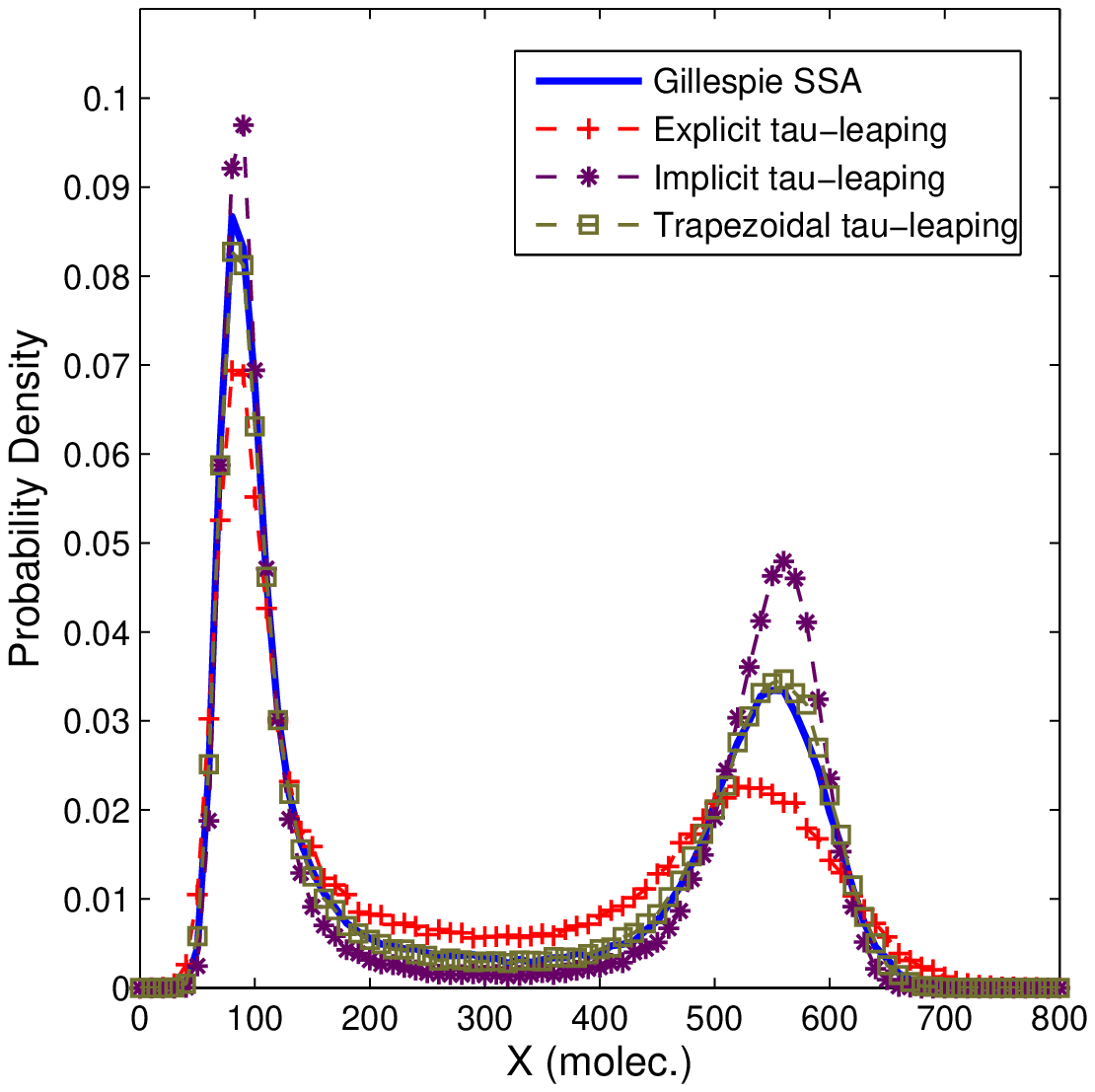}
  }
  \subfigure[Histograms obtained with Gillespie's SSA,
  the new fully implicit methods, and the new implicit order two weak
  Taylor tau-leaping methods.]{
    \includegraphics[width=3.1in]{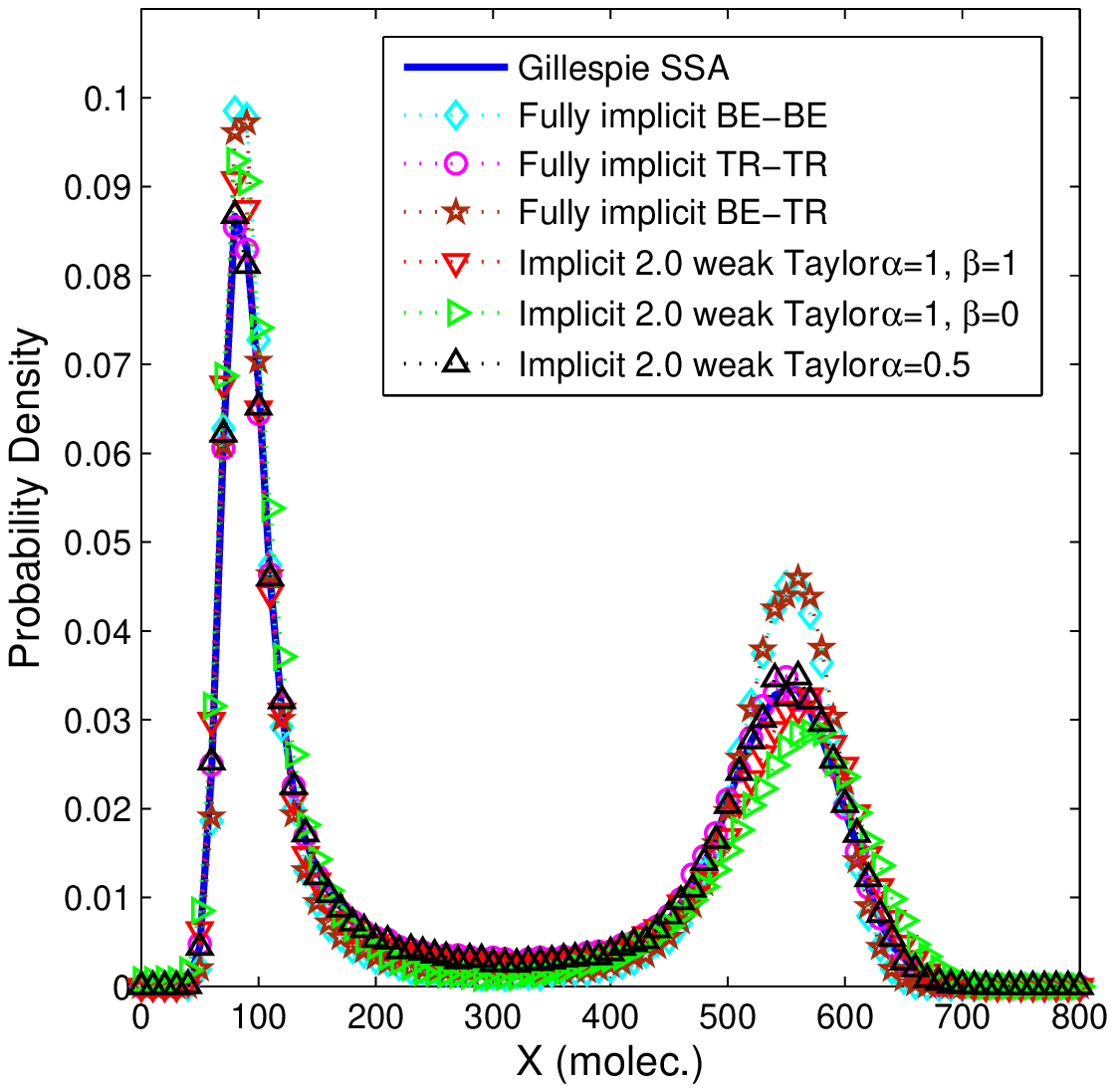}
  }
  \caption{The histograms of the number of molecules $X$ at the final time for
  the Schl\"{o}gl bistable system~\eqref{eqn:schlogl_system}.
  All histograms are based on 100,000 runs of the corresponding methods with a fixed stepsize $\tau=0.4$ seconds.}
  \label{fig:Schlogl_hist}
\end{figure}

The histograms generated from 100,000 independent samples of SSA,
existing improved SSA methods, and proposed methods including fully
implicit tau-leaping methods and implicit order two weak Taylor
methods with fixed stepsize $\tau=0.4$ are shown in
Figure~\ref{fig:Schlogl_hist}. We notice that the histogram given by
the trapezoidal tau-leaping method, fully implicit TR--TR method,
and implicit order two weak Taylor method with $\alpha=0.5$ are very
close to the exact SSA method than other methods for the specific
time step as the histogram of the decaying-dimerizing system. The
histograms produced by the fully implicit BE--BE and BE--TR exhibit
damping effect (sharp peaks) while the histogram given by the
implicit order two weak Taylor method with $\alpha=1.0$, $\beta=1.0$
and $\alpha=1.0$, $\beta=0.0$ methods provoke a little wide varying
results (broad peaks).

\begin{table*}[t!]
\caption{The mean, variance, distance, and elapsed CPU times (in
seconds) for $X$ at $T=4$ based on 100,000 samples for different
stepsizes of the Schl\"{o}gl bistable
system~\eqref{eqn:schlogl_system}.}
\vskip 0.10in
\centering
\linespread{1.00}
\footnotesize
\begin{tabular}{|C{2.3cm}|C{1.8cm}|C{2.6cm}|C{2.4cm}|C{2.4cm}| C{2.4cm} |} \hline
\multicolumn{2}{|c|}{} & \multicolumn{4}{|c|}{\textbf{Stepsize
($\tau$ in seconds)}} \\ \hline
\multicolumn{1}{|c|}{\textbf{Method}} &
\multicolumn{1}{|c|}{\textbf{Metrics}} &
  \multicolumn{1}{|c|}{$0.8$} &
  \multicolumn{1}{|c|}{$0.4$} &
  \multicolumn{1}{|c|}{$0.2$} &
  \multicolumn{1}{|c|}{$0.1$} \\ \hline \hline
\multicolumn{1}{|c|}{Gillespie} & \multicolumn{1}{|c|}{Mean (Var)} & \multicolumn{4}{|c|}{305.2 (46465.9)} \\
\multicolumn{1}{|c|}{SSA} & \multicolumn{1}{|c|}{CPU time} &
\multicolumn{4}{|c|}{682.96} \\ \hline
Explicit       & Mean (Var)     & 296.9 (40957.6) & 306.2 (42915.6) & 309.5 (44981.5) & 308.5 (45929.9) \\
tau-leaping    & Distance       & 5.680 & 3.155 & 2.057 & 1.860 \\
               & CPU time       & 1.41 & 2.1 & 3.43 & 6.21 \\ \hline
Implicit       & Mean (Var)     & 343.4 (52245.0) & 326.3 (49876.8) & 316.9 (48364.8) & 315.1 (47644.7) \\
tau-leaping    & Distance       & 4.464 & 2.877 & 2.136 &  1.936 \\
               & CPU time       & 4.41 & 7.03 & 12.24 & 22.4 \\ \hline
Trapezoidal    & Mean (Var)     & 324.6 (47837.6) & 317.4 (47161.6) & 312.6 (46727.0) & 311.2 (46719.1) \\
tau-leaping    & Distance       & 2.036 & 1.906 & 1.849 & 1.818 \\
               & CPU time       & 4.2 & 6.79  & 12.07 & 22.6 \\ \hline \hline
Fully implicit & Mean (Var)     & 316.4 (51137.7) & 318.8 (49359.6) & 313.5 (47919.2) & 312.2 (47401.1) \\
BE--BE         & Distance       & 4.360 & 2.808 & 2.158 & 1.956 \\
               & CPU time       & 8.64 & 13.6 & 23.74 & 43.86 \\ \hline
Fully implicit & Mean (Var)     & 316.2 (47195.7) & 312.4 (46743.9) & 312.2 (46624.0) & 309.9 (46601.9) \\
TR--TR         & Distance       & 1.943 & 1.857 & 1.836 & 1.818 \\
               & CPU time       & 8.13 & 13.63 & 24.51 & 46.4 \\ \hline
Fully implicit & Mean (Var)     & 335.5 (51920.4) & 322.3 (49566.8) & 315.9 (48011.9) & 311.1 (47325.1) \\
BE--TR         & Distance       & 4.417 & 2.761 & 2.147 & 1.917 \\
               & CPU time       & 8.80 & 13.38 & 24.98 & 46.76 \\ \hline\hline
Implicit 2.0   & Mean (Var)     & 1122.4 (51112.5) & 310.3 (49157.7) & 310.2 (47332.8) & 310.0 (46612.9) \\
weak Taylor    & Distance       & 3.501 & 1.890 & 1.830 & 1.766 \\
($\alpha=1,\beta=1$) & CPU time & 12.53 & 18.72 & 30.98 & 55.08 \\
\hline
Implicit 2.0   & Mean (Var)     & 296.4 (50810.1) & 306.2 (46870.6) & 309.5 (46566.0) & 309.7 (46498.5) \\
weak Taylor    & Distance       & 2.475 & 1.869 & 1.842 & 1.839 \\
($\alpha=1,\beta=0$) & CPU time & 11.74 & 17.48 & 28.76 & 52.64 \\
\hline
Implicit 2.0   & Mean (Var)     & 313.2 (47441.4) & 309.9 (46880.3) & 309.7 (46494.3) & 310.2 (46503.7) \\
weak Taylor    & Distance       & 1.862 & 1.840 &  1.809 & 1.803 \\
($\alpha=0.5$) & CPU time       & 10.71 & 16.34 &  26.47 & 50.23 \\
\hline
\end{tabular}
\label{table:schlogl_compare}
\end{table*}

Table~\ref{table:schlogl_compare} shows the mean, variance,
distance, and elapsed CPU times based on 100,000 samples generated
by different methods for fixed stepsizes. Four fixed stepsizes
$\tau=0.8/k$ where $k=1,2,4,8$ were selected to evaluate accuracy
for each time step. The variance for all methods are large for the
bistability property of the system. Proposed fully implicit TR--TR,
and the implicit second order weak Taylor with $\alpha=0.5$ produce
accurate results even with large stepsize $\tau=0.8$.

\begin{figure}[h!]
  \begin{center}
  \center
  \scalebox{0.7}{\includegraphics{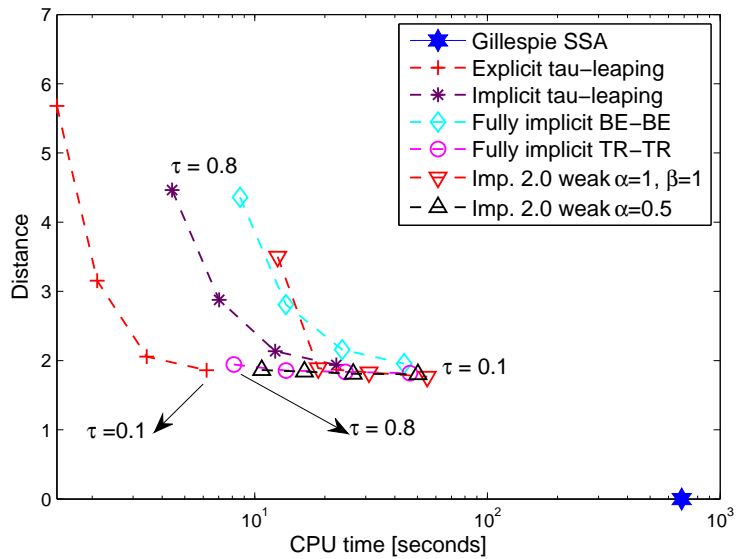}}
  \caption{Relationship between solution accuracy measured by the distribution distance \eqref{eq:dist}
  and CPU time for different methods applied to the Schl\"{o}gl bistable
  system~\eqref{eqn:schlogl_system}.}
  \label{fig:Schlogl_dist_cpu}
  \end{center}
\end{figure}

Figure~\ref{fig:Schlogl_dist_cpu} shows the relationship between
distance of two distributions (the SSA and each accelerated method
distributions) and computation time for the different stepsizes of
Schl\"{o}gl bistable system. As the previous dimer reaction system,
the fully implicit TR--TR and the implicit second order weak Taylor
method with the $\alpha=0.5$ show small distance (good accuracy)
compared to other accelerated methods with the big stepsize
$\tau=0.8$. 100,000 simulations of the fully implicit TR--TR method
with the $\tau=0.8$ took 8.13 seconds with accuracy. With the
limited results investigated here, the explicit tau-leaping method
is the most efficient for this system. 100,000 simulations of the
explicit tau-leaping method for the small stepsize $\tau=0.1$ took
6.21 seconds with small distance as ones of fully implicit TR--TR
results for the stepsize $\tau=0.4$. All accelerated methods show
efficiency (at least 10 times faster) compared to the SSA that took
683 seconds for 100,000 simulations.

\subsection{The ELF System}
We now consider a more complex system containing 8 species and 12
reactions~\cite{Elf04, Marquez07, Hu11} to evaluate the accuracy of
the proposed tau-leaping methods. We use the initial conditions and
parameter values given in the literature \cite{Hu11}. The chemical
reactions, propensity functions, and initial values are listed in
Table \ref{table:comp_system}.

\begin{table}[b]
\caption{List of reactions and propensity functions for the ELF
system.}
\vskip 0.10in
\centering
\linespread{1.0}
\footnotesize
\begin{tabular}{|p{0.6cm}p{3.2cm}p{2.9cm}l|p{0.6cm}p{1.3cm}p{2.2cm}|} \hline\hline
& \textbf{Reaction} & \textbf{Propensity} & \textbf{Rate constant} &
& \textbf{Species} & \textbf{Initial value}\\ \hline
$R_1$  & $E_A \rightarrow E_A + A$            & $a_1  = c_1[E_A]$         & $c_1  = 15$     & $X_1$ & $A$ & 2000 molec.\\
$R_2$  & $E_B \rightarrow E_B + B$            & $a_2  = c_2[E_B]$         & $c_2  = 15$     & $X_2$ & $B$ & 1500 molec.\\
$R_3$  & $E_A + B \rightarrow E_{A}B$         & $a_3  = c_3[E_A][B]$      & $c_3  = 0.0001$ & $X_3$ & $E_A$ & 950 molec.\\
$R_4$  & $E_{A}B \rightarrow E_A + B$         & $a_4  = c_4[E_{A}B]$      & $c_4  = 0.6$    & $X_4$ & $E_B$ & 950 molec.\\
$R_5$  & $E_{A}B + B \rightarrow E_{A}B_{2}$  & $a_5  = c_5[E_{A}B][B]$   & $c_5  = 0.0001$ & $X_5$ & $E_{A}B$ & 200 molec.\\
$R_6$  & $E_{A}B_{2} \rightarrow E_{A}B + B$  & $a_6  = c_6[E_{A}B_{2}]$  & $c_6  = 0.6$    & $X_6$ & $E_{A}B_{2}$ & 50  molec.\\
$R_7$  & $A \rightarrow 0$                    & $a_7  = c_7[A]$           & $c_7  = 0.5$    & $X_7$ & $E_{B}A$ & 200 molec.\\
$R_8$  & $E_B + A \rightarrow E_{B}A$         & $a_8  = c_8[E_B][A]$      & $c_8  = 0.0001$ & $X_8$ & $E_{B}A_{2}$ & 50 molec.\\
$R_9$  & $E_{B}A \rightarrow E_B + A$         & $a_9  = c_9[E_{B}A]$      & $c_9  = 0.6$ &&& \\
$R_{10}$ & $E_{B}A + A \rightarrow E_{B}A_{2}$  & $a_10 = c_10[E_{B}A][A]$  & $c_10 = 0.0001$ &&&\\
$R_{11}$ & $E_{B}A_{2} \rightarrow E_{B}A + A$  & $a_11 = c_11[E_{B}A_{2}]$ & $c_11 = 0.6$ &&&\\
$R_{12}$ & $B \rightarrow 0$                    & $a_12 = c_12[B]$ &
$c_12 = 0.5$&&& \\ \hline\hline
\end{tabular}
\label{table:comp_system}
\end{table}

\begin{figure}[!h]
  \centering
  \subfigure[ $~~\tau=0.04$ sec.]{
    \includegraphics[width=3in]{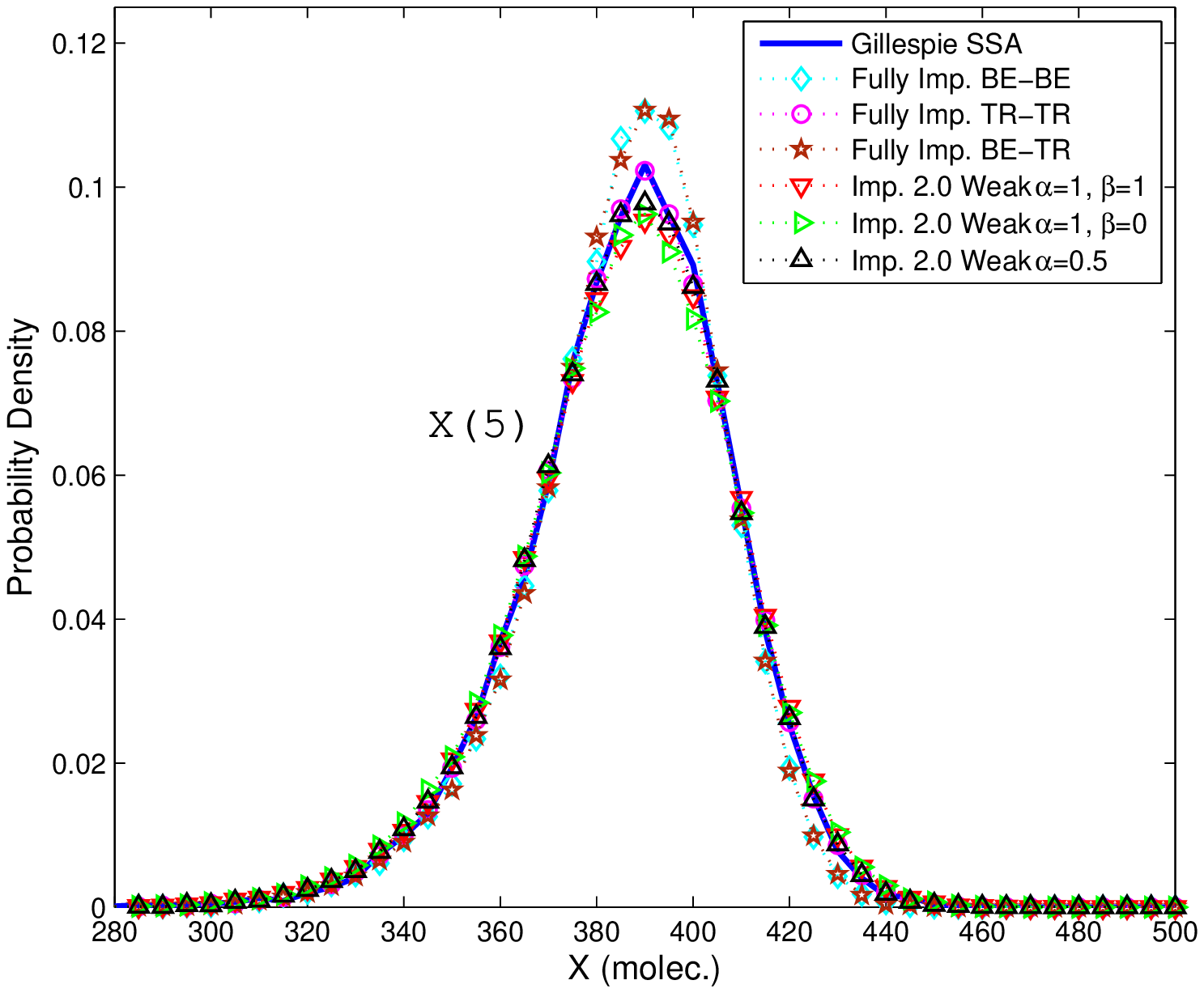}
  }
  \subfigure[ $~~\tau=0.005$ sec.]{
    \includegraphics[width=3in]{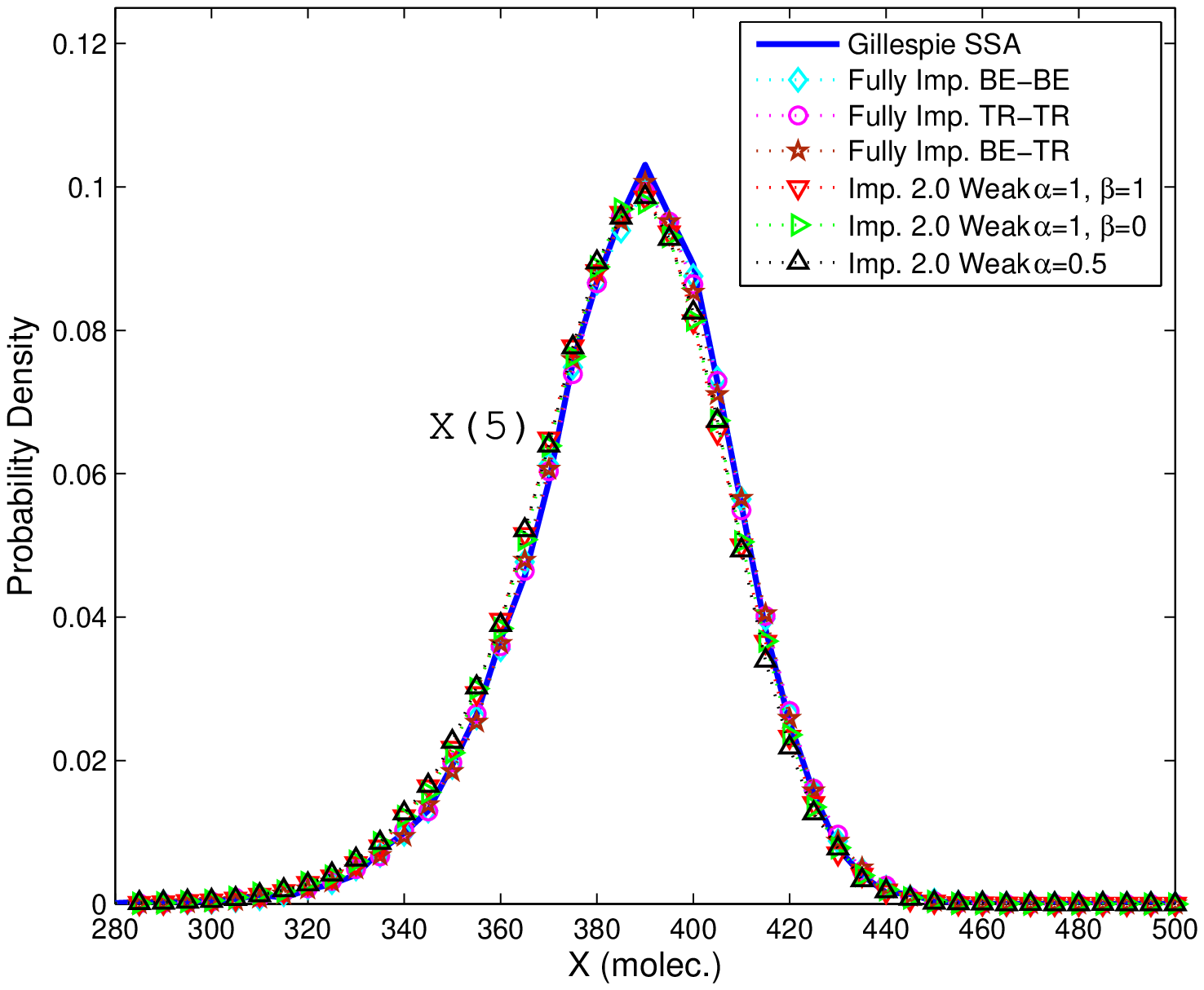}
  }
  \caption{The histograms of $X_5$ at the final time obtained with different, fixed stepsizes for the ELF system (Table
\ref{table:comp_system}). Each histogram uses 100,000 samples.}
  \label{fig:system3_X5_hist}
\end{figure}

\begin{figure}[!h]
  \centering
  \subfigure[$~~\tau=0.04$ sec.]{
    \includegraphics[width=3in]{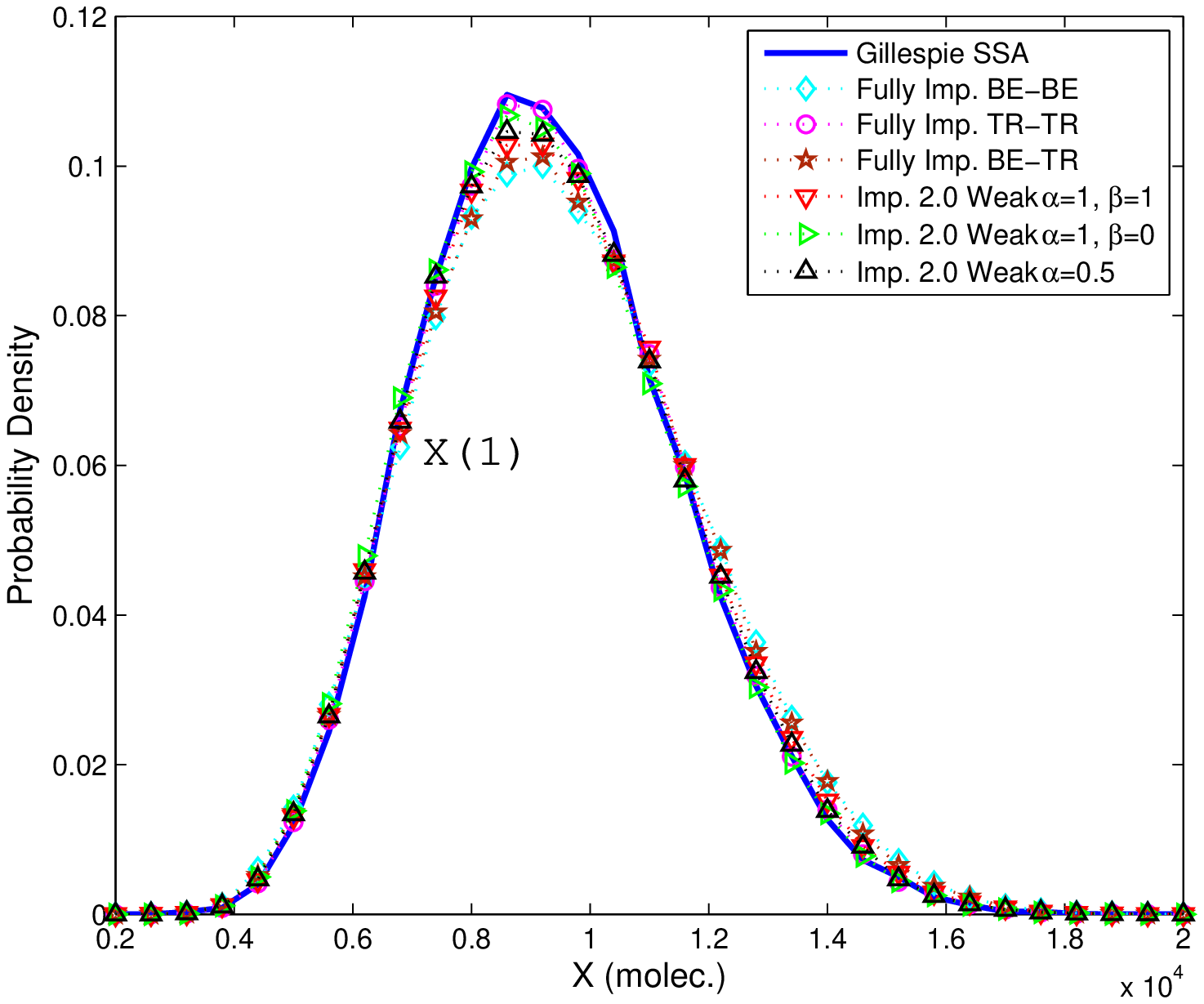}
  }
  \subfigure[$~~\tau=0.005$ sec.]{
    \includegraphics[width=3in]{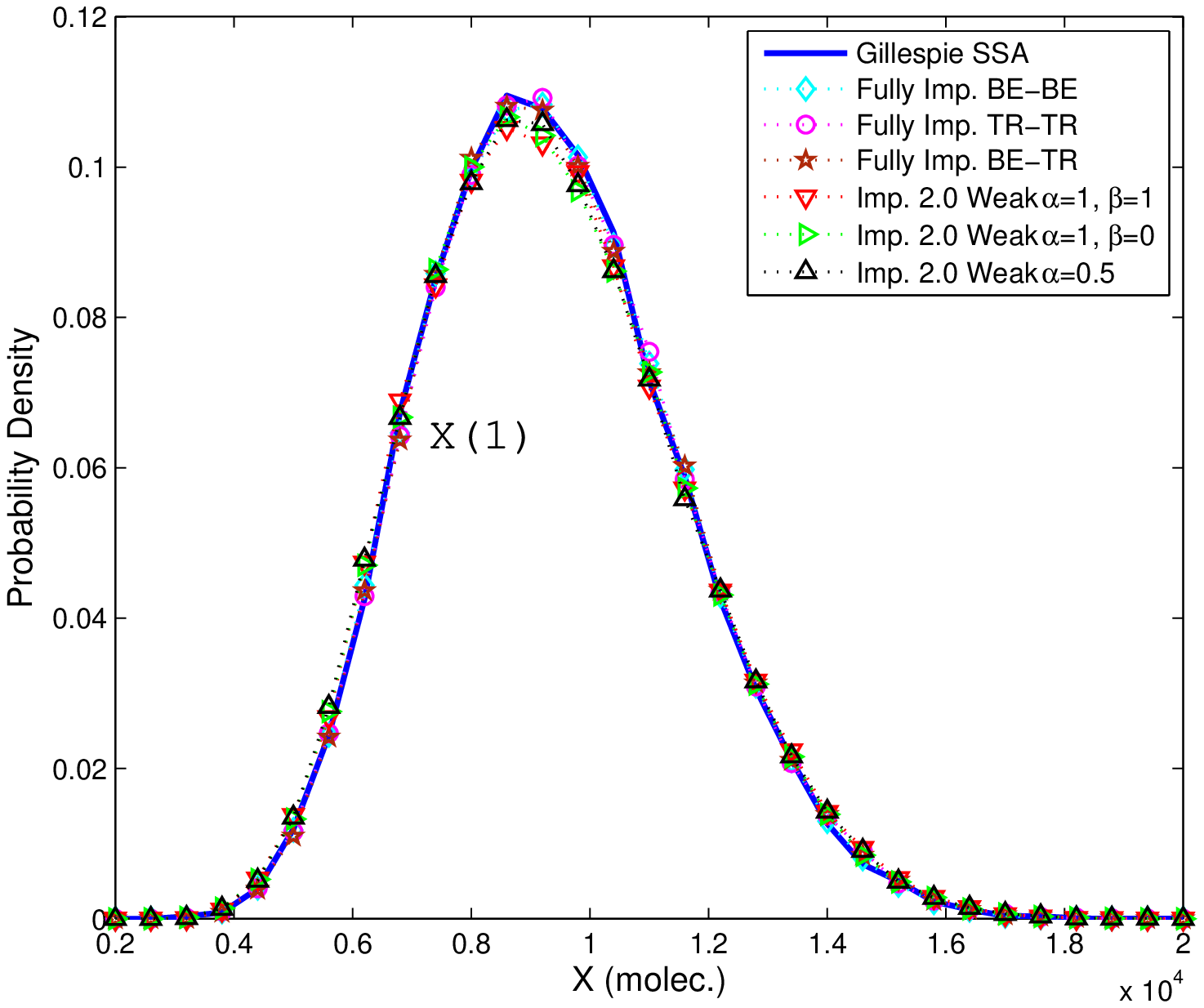}
  }
  \caption{The histograms of $X_1$ at the final time obtained with different, fixed stepsizes for the ELF system (Table
\ref{table:comp_system}). Each histogram uses 100,000 samples.}
  \label{fig:system3_X1_hist}
\end{figure}

We consider the simulation time interval $[0, 3]$ seconds, and
perform 100,000 independent runs with the Gillespie SSA and with
each one of the accelerated methods. The histograms of  $X_5$ and
$X_1$ concentrations at the final time are presented in Figures
\ref{fig:system3_X5_hist} and \ref{fig:system3_X1_hist},
respectively, for different fixed time steps between $\tau=0.04$ and
$\tau=0.005$ seconds. Figure \ref{fig:system3_X5_hist} shows a
similar qualitative behavior as in the previous stiff examples. For
a large stepsize $\tau=0.04$ seconds, the histograms produced by the
fully implicit BE--BE and BE--TR methods exhibit a weak damping
effect (small sharp peaks), while the histograms given by the
implicit order two weak Taylor methods with $\alpha=1.0$ exhibit a
dispersive effect (broader peaks). Figure \ref{fig:system3_X1_hist}
shows a different behavior. For a large stepsize $\tau=0.04$ seconds
the BE--BE, the BE--TR, and the implicit order 2.0 weak Taylor with
$\alpha=1.0$ methods show dispersive behavior (broad peaks).
Therefore the errors in variance for the ELF system have a complex
behavior when stepsizes are very large. In Figures
\ref{fig:system3_X5_hist} and \ref{fig:system3_X1_hist}, the
histograms given by the fully implicit TR--TR method and implicit
order two weak Taylor method with $\alpha=0.5$ are very similar to
the exact SSA histogram. If the stepsize $\tau$ is decreased to
$\tau=0.005$ seconds, all approximation methods show very good
accuracy.

\begin{figure}[!t]
  \centering
  \subfigure[Error in the distribution of $X_5$.]{
    \includegraphics[width=3.1in]{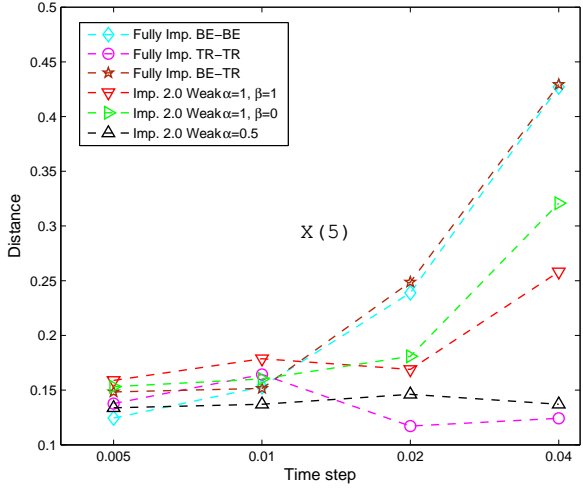}
  }
    \subfigure[Error in the distribution of  $X_1$.]{
    \includegraphics[width=3.1in]{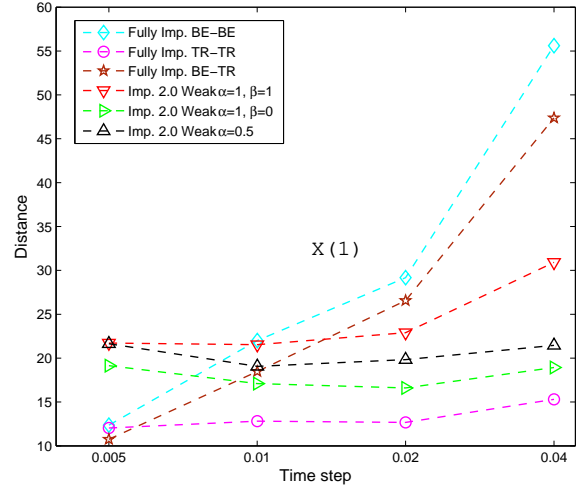}
  }
  \caption{The relationship between the error in distribution
  (the distance \eqref{eq:dist} between SSA and each of the proposed methods' histograms) and the different
  stepsizes for $X_5$ and $X_1$ for the ELF system.}
  \label{fig:system3_dist}
\end{figure}

\begin{figure}[h!]
  \begin{center}
  \center
  \scalebox{0.65}{\includegraphics{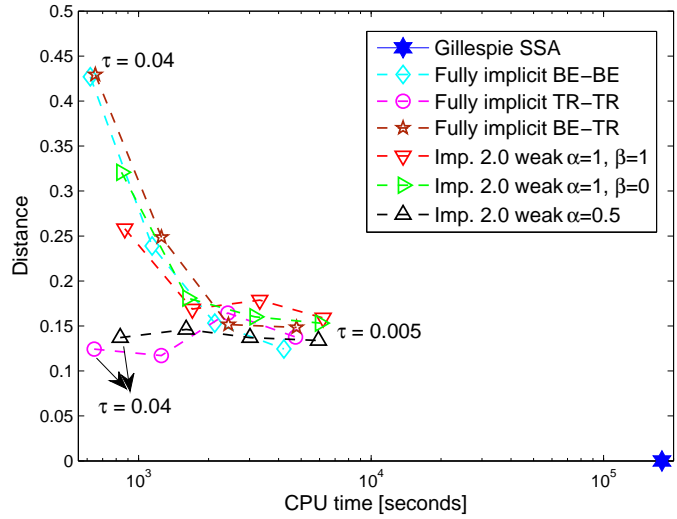}}
  \caption{The relationship between accuracy and CPU time for
  $X_5$ of the ELF system}
  \label{fig:system3_dist_cpu}
  \end{center}
\end{figure}

Figures~\ref{fig:system3_dist} (a) and (b) show the error in
distribution (the distance \eqref{eq:dist} between the SSA and each
of the accelerated methods' histograms) versus simulation stepsize
for the ELF system. The y-scale  in Figure~\ref{fig:system3_dist}
(b) is much larger than that of Figure~\ref{fig:system3_dist} (a)
because the number of molecules for $X_1$ is much larger than that
of $X_5$ (see the Figures \ref{fig:system3_X5_hist} and
\ref{fig:system3_X1_hist}). The results indicate that, similar to
the previous examples, the TR--TR and the implicit second order weak
Taylor method with the $\alpha=0.5$ are the most accurate
accelerated methods.

Figure~\ref{fig:system3_dist_cpu} shows the relationship between
accuracy and CPU time for the different stepsizes of the ELF system.
The accuracy is measured by the distance~\eqref{eq:dist} between the
accelerated method and the SSA histograms for $X_5$, as in
Figure~\ref{fig:system3_dist} (a). 100,000 simulation of the SSA
took 178,364 seconds (approximately 50 hours), while 100,000
simulations of the implicit order two weak Taylor method with
$\alpha=1.0$ and $\beta=1.0$ for the smallest stepsize $\tau=0.005$
took 6,216 seconds (3.5\% of the SSA time) and provided an accurate
solution (distance value is only 0.15). For the largest fixed
stepsize $\tau=0.04$ seconds, the fully implicit TR--TR and the
implicit second order weak Taylor method with the $\alpha=0.5$
provide high accuracy and high efficiency (only 0.4\% of the SSA
time).

\section{Conclusions} \label{sec:conclusions}
This paper develops new implicit tau-leaping-like algorithms for the
solution of stochastic chemical kinetic systems. The fully implicit
tau-leaping methods, ``BE--BE'', ``TR--TR'', and ``BE--TR'', are
motivated by the fact that existing implicit tau-leaping algorithms
treat implicitly only the mean part of the Poisson process. The
newly proposed methods treat implicitly the variance of the Poisson
variables as well. The implicit second order weak Taylor tau-leaping
methods are motivated by the theory of weakly convergent
discretizations of stochastic differential equations, and by the
fact that Poisson variables with large mean are well approximated by
normal variables.

Theoretical stability and consistency analyses are carried out on a
standard test problem -- the reversible isomerization reaction. The
fully implicit tau-leaping methods are unconditionally stable; the
implicit second order weak Taylor tau-leaping methods with
$\alpha=1.0$ are conditionally stable, and with $\alpha=0.5$
unconditionally stable. The asymptotic means of the solutions given
by all proposed methods converge to the analytical mean of the test
problem. The asymptotic variances of the proposed methods, however,
converge to different values, as it is also the case for traditional
tau-leaping methods.

Numerical experiments are carried out using the decaying-dimerizing
system, the bistable Schl\"{o}gl reaction system, and the ELF system
to validate the theoretical results. The accuracy of the solutions
is evaluated by comparing the probability densities obtained with
the new methods and with Gillespie's SSA. The numerical results
verify that the prosed methods are accurate, with an efficiency
comparable to that of the traditional implicit tau-leaping methods.
The theoretical analyses and numerical experiments shows that the
fully implicit TR--TR and the implicit second order weak Taylor
tau-leaping methods with $\alpha=0.5$ are the most accurate methods
for large stepsizes.

\section*{Acknowledgements}
This work was supported in part by awards NIGMS/NIH 5 R01 GM078989,
NSF CMMI--1130667, NSF CCF-0916493, OCI-0904397, NSF DMS--0915047,
NSF CCF--1218454, AFOSR 12-2640-06, and by the Computational Science
Laboratory at Virginia Tech.


\bibliographystyle{elsarticle-num}
\bibliography{NewImplicitSSA}

\begin{thebibliography}{10}
\expandafter\ifx\csname url\endcsname\relax
  \def\url#1{\texttt{#1}}\fi
\expandafter\ifx\csname urlprefix\endcsname\relax\def\urlprefix{URL }\fi
\expandafter\ifx\csname href\endcsname\relax
  \def\href#1#2{#2} \def\path#1{#1}\fi

\bibitem{McAdams97}
H.~H. McAdams, A.~Arkin, Stochastic mechanisms in gene expression, Proc. Natl.
  Acad. Sci. 94~(3) (1997) 814--819.

\bibitem{Gill92}
D.~T. Gillespie, A rigorous derivation of the chemical master equation, Physica
  A 188~(1--3) (1992) 404--425.

\bibitem{Kampen07}
N.~G. van Kampen, Stochastic Processes in Physics and Chemistry, North Holland,
  North Holland, Netherlands.

\bibitem{Gill77}
D.~T. Gillespie, Exact stochastic simulation of coupled chemical reactions,
  Journal of Physical Chemistry 81~(25) (1977) 2340--2361.

\bibitem{Gill01}
D.~T. Gillespie, Approximate accelerated stochastic simulation of chemically
  reacting systems, Journal of Chemical Physics 115~(4) (2001) 1716--1733.

\bibitem{Rath03}
M.~Rathinam, L.~R. Petzold, Y.~Cao, D.~T. Gillespie, Stiffness in stochastic
  chemically reacting systems: The implicit tau-leaping method, Journal of
  Chemical Physics 119~(24) (2003) 12784--12794.

\bibitem{Cao05}
Y.~Cao, L.~Petzold, Trapezoidal tau-leaping formula for the stochastic
  simulation of biochemical systems, in: Proceedings of Foundations of Systems
  Biology in Engineering (FOSBE 2005), 2005, pp. 149--152.

\bibitem{Cao04}
Y.~Cao, H.~Li, L.~Petzold, Efficient formulation of the stochastic simulation
  algorithm for chemically reacting systems, Journal of Chemical Physics
  121~(9) (2004) 4059--4067.

\bibitem{Gill03}
D.~T. Gillespie, L.~R. Petzold, Improved leap-size selection for accelerated
  stochastic simulation, Journal of Chemical Physics 119~(16) (2003)
  8229--8234.

\bibitem{Rath05}
M.~Rathinam, L.~R. Petzold, Y.~Cao, D.~T. Gillespie, Consistency and stability
  of tau leaping schemes for chemical reaction systems, SIAM Journal of
  Multiscale Modeling and Simulation 4~(3) (2005) 867--895.

\bibitem{Tian01}
T.~Tian, K.~Burrage, Implicit taylor methods for stiff stochastic differential
  equations, Applied Numerical Mathematics 38~(1-2) (2001) 167--185.

\bibitem{Li07}
T.~Li, Analysis of explicit tau-leaping schemes for simulating chemically
  reacting systems, Multiscale Modeling and Simulation 6~(2) (2007) 417--436.

\bibitem{Hu11}
Y.~Hu, T.~Li, B.~Min, A weak second order tau-leaping method for chemical
  kinetic systems, Journal of Chemical Physics 135~(2) (2011) 024113.

\bibitem{Kloeden99}
P.~E. Kloeden, E.~Platen, Numerical Solution of Stochastic Differential
  Equations, Springer, New York, NY.

\bibitem{Cao04-2}
Y.~Cao, L.~R. Petzold, M.~Rathinam, D.~T. Gillespie, The numerical stability of
  leaping methods for stochastic simulation of chemically reacting systems,
  Journal of Chemical Physics 121~(24) (2004) 12169--12178.

\bibitem{Ahn11-HPC11}
T.-H. Ahn, A.~Sandu, Fully implicit tau-leaping methods for the stochastic
  simulation of chemical kinetics, in: Proceedings of the 2011 Spring
  Simulation Multiconference, SpringSim '11, Society for Computer Simulation
  International, Boston, MA, USA, 2011.

\bibitem{Ahn11-ICCS11}
T.-H. Ahn, A.~Sandu, Implicit second order weak taylor tau-leaping methods for
  the stochastic simulation of chemical kinetics, Vol.~4, 2011, pp. 2297 --
  2306, proceedings of the International Conference on Computational Science,
  ICCS 2011.

\bibitem{Gikhman72}
I.~I. Gikhman, A.~V. Skorokhod, Stochastic Differential Equations, Springer,
  New York, NY, 1972.

\bibitem{Ross06}
S.~M. Ross, Introduction to Probability Models, Ninth Edition, Academic Press,
  Inc., Orlando, FL, USA, 2006.

\bibitem{Emmert08}
F.~Emmert-Streib, M.~Dehmer, Information Theory and Statistical Learning,
  Springer, New York, NY, 2008.

\bibitem{Elf04}
J.~Elf, M.~Ehrenberg, Spontaneous separation of bi-stable biochemical systems
  into spatial domains of opposite phases, Systems Biology, IEE Proceedings
  1~(2) (2004) 230 -- 236.

\bibitem{Marquez07}
T.~T. Marquez-Lago, K.~Burrage, Binomial tau-leap spatial stochastic simulation
  algorithm for applications in chemical kinetics, Journal of Chemical Physics
  127~(10) (2007) 104101.

\end{thebibliography}







\end{document}